\documentclass{elsart}
\usepackage{graphicx,epsfig,wrapfig,lineno}
\begin{document}


\newpage
\begin{frontmatter}

\title{First Year Performance of The IceCube Neutrino Telescope}

\begin{center}
\vskip 0.5 cm
{

 \collab{IceCube Collaboration}
\author[Utrecht]{A.~Achterberg},
\author[Zeuthen]{M.~Ackermann},
\author[Christchurch]{J.~Adams},
\author[Mainz]{J.~Ahrens},
\author[Madison]{K.~Andeen},
\author[PennPhys]{D.~W.~Atlee},
\author[Madison]{J.~Baccus},
\author[Princeton]{J.~N.~Bahcall\thanksref{dec}},
\author[Bartol]{X.~Bai},
\author[BrusselsVrije]{B.~Baret},
% ...
\address[BrusselsVrije]{Vrije Universiteit Brussel, Dienst ELEM, B-1050 Brussels, Belgium}
% ...
\address[Christchurch]{Dept.~of Physics and Astronomy, University of Canterbury, Private Bag 4800, Christchurch, New Zealand}
% ...
\address[Madison]{Dept.~of Physics, University of Wisconsin, Madison, WI 53706, USA}
\address[Mainz]{Institute of Physics, University of Mainz, Staudinger Weg 7, D-55099 Mainz, Germany}
% ...
\address[Bartol]{Bartol Research Institute, University of Delaware, Newark, DE 19716, USA}
% ...
\address[Princeton]{Institute for Advanced Study, Princeton, NJ 08540, USA}
% ...
\address[PennPhys]{Dept.~of Physics, Pennsylvania State University, University Park, PA 16802, USA}
% ...
\address[Utrecht]{Dept.~of Physics and Astronomy, Utrecht University/SRON, NL-3584 CC Utrecht, The Netherlands}
% ...
\address[Zeuthen]{DESY, D-15735 Zeuthen, Germany}
\thanks[dec]{Deceased}
\thanks[Bari]{on leave of absence from Universit\`a di Bari, Dipartimento di Fisica, I-70126, Bari, Italy}
\thanks[Kalmar]{affiliated with Dept.~of Chemistry and Biomedical Sciences, Kalmar University, S-39182 Kalmar, Sweden}
\thanks[Valencia]{affiliated with IFIC (CSIC-Universitat de Val\`encia), A. C. 22085, 46071 Valencia, Spain}
}
\end{center} 

\newpage

\begin{abstract}
The first sensors of the IceCube neutrino observatory were deployed at 
the South Pole 
during the austral summer of 2004-05 and have been producing data 
since February 2005.  
One string of 60 sensors buried in the ice and a surface array of 8 ice Cherenkov 
tanks took data until December 2005 when deployment of the next set of
strings and tanks began.  We have analyzed these data, 
demonstrating that the
performance of the system meets or exceeds design requirements.  Times
are determined across the whole array to a relative precision of better than 3
nanoseconds, allowing reconstruction of muon tracks and light bursts
in the ice, of air-showers in the surface array and of events seen in
coincidence by surface and deep-ice detectors separated by up to 2.5 km.
\end{abstract}
\end{frontmatter}

\newpage
\section{Introduction}
\label{section:introduction}
Concepts for building a detector large enough to study neutrinos from
astrophysical sources have evolved since the early days of
experimental neutrino physics nearly 50 years
ago~\cite{Greisen,Markov}.  First successes occurred with the
observations of neutrinos from SN1987A \cite{Kamiokande,IMB,Baksan} 
and of solar neutrinos \cite{Homestake,Sage,Gallex,KamSolar,SKsolar,SNO}. 
Efforts to build a detector large enough to identify
the less abundant but higher energy neutrinos produced by hadronic
interactions in cosmic-ray sources began with DUMAND~\cite{DUMAND}.
Neutrino telescopes currently in operation are NT200+ (Baikal)~\cite{Baikal}
and AMANDA~\cite{AMANDA}.  These detectors work by observing a large
volume of clear water or ice to detect Cherenkov light from 
relativistic charged
particles produced in neutrino interactions in or near the target
volume.

Several arguments lead to the conclusion that a fiducial volume of
at least a cubic kilometer is needed to observe neutrinos from high-energy astrophysical sources~\cite{kilometer}. Several groups are exploring
use of optical Cherenkov radiation in water or ice to detect
neutrinos~\cite{Antares,Nemo,Nestor,IceCube}.  
Neutrino telescopes deep in the ice can be 
calibrated using a surface air-shower array, that can improve
the rejection of the background associated with cosmic-rays, and allow 
the study of cosmic-rays.
Projects based on other techniques (horizontal air-showers, radio and
acoustic signals from neutrino-induced events) are also being
developed~\cite{review}.  The latter generally have energy thresholds several
orders of magnitude higher than the 50-100 GeV range of ice and water Cherenkov detectors.

IceCube builds on the successful deployment and operation since 1996
of the AMANDA neutrino telescope ~\cite{icrc}.  AMANDA consists of 677
optical sensors distributed on 19 strings instrumenting a volume of
more than $10^7$ m$^3$ at a depth between 1500 and 2000 meters in the ice at the South Pole.  In addition to its larger volume, IceCube differs from
AMANDA in two significant ways.  First, in the IceCube sensors, the signals are digitized in
the optical sensors to minimize loss of information from degradation
of analog signals sent over long distances.  
Second, a free-running 20 MHz oscillator in each IceCube sensor serves as a local clock and provides time stamps (at 20 MHz and 40 MHz) for internal operations in the sensor, including timing the arrival of photons.  
The clock drift is less than 2 ns per second. This local clock is calibrated automatically relative to a master clock on the surface.  All time-stamps are converted to Universal time (UTC) at the central counting station.

The IceCube neutrino observatory will consist of 4800 optical sensors
or Digital Optical Modules (DOMs) installed on 80 strings between 1450 m 
and 2450 m below the surface \cite{design}.  The In-Ice array is complemented by a surface array, called IceTop. IceTop will
consist of 160 ice-tanks, in pairs, near the top of each string. Each 
tank has two DOMs for redundancy and extended dynamic range.

This paper describes the performance of the first string and eight 
surface tanks that were installed between November 2004 and January  
2005.  Section \ref{section:DetectorOverview} introduces the design, 
production, deployment and configuration of the detector.  Detailed 
descriptions of the digital optical module and its electronics, the 
timing method, the surface detectors and data acquisition will be 
given in separate papers.  After the overview, we describe the
calibration of gain and timing (section \ref{section:Calibration}).
Then follow three sections that deal in 
turn with reconstruction of muons in the deep string, reconstruction 
of air-showers with IceTop and analysis of coincidences with 
externally triggered detectors, including AMANDA.  We summarize 
results of the first season in a concluding section and describe the 
plan for completion of the IceCube construction project. 
\section{Overview of the Detector}
\label{section:DetectorOverview}
The design of IceCube calls for 80 strings each instrumented with 60
DOMs capable of detecting signals over a wide dynamic range, from a single 
photon to several thousands arriving within a few microseconds of 
each other~\cite{design}.  
Strings will be deployed in a triangular grid pattern with a characteristic 
spacing of 125 m enclosing an area 
of $1$~km$^2$. 
Each hole cable, which carries 60 DOMs, is connected to a surface junction box  
placed between the two IceTop tanks. The IceTop DOMs are also connected to the 
surface junction box.  A cable from the surface junction box to the central 
counting house carries all DOM cables and service wires.  
Signals are digitized and time 
stamped in the modules. Times and waveforms from several modules are used to 
reconstruct events from the Cherenkov light emitted by charged particles in
the deep ice and in the IceTop tanks.

\subsection{The Digital Optical Module (DOM)} 
\label{section:DOM}
\begin{figure}[htb]
\begin{center}
\includegraphics[width=10cm]{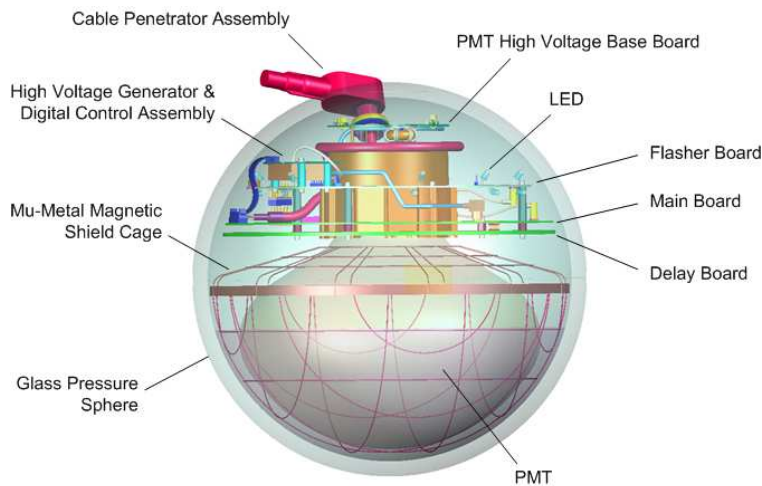}
\end{center}
\caption{A schematic view of an IceCube Digital Optical Module. }
\label{ADOM}
\end{figure}

The Digital Optical Module is the fundamental detector element
of IceCube.  It consists of a 25 cm diameter R7081-02 Hamamatsu Photo-multiplier Tube (PMT) and a suite
of electronics board assemblies contained within a 35.6 cm-diameter
glass pressure housing (Fig. \ref{ADOM}).  The DOM achieves high
accuracy and a wide dynamic range by internally digitizing and 
time-stamping the photonic signals and transmitting packetized 
digital data to the surface.  This scheme, which was first demonstrated in 
AMANDA String-18~\cite{DOM1}, allows each DOM to operate as a complete and
autonomous data acquisition system.  Each DOM can independently perform 
functions such as PMT gain calibration, time calibration with the 
master clock system (section \ref{section:Calibration}), data 
packaging and response to remote commands for change of configuration.

Since the DOMs are inaccessible once deployed in the deep ice, they have been 
designed to operate reliably for 15 years in a cold, high-pressure 
environment.  High-reliability commercial parts were used where 
possible, and all sub-assemblies underwent a rigorous screening process 
involving functional and environmental stress testing, and/or 
inspection before being integrated into a DOM.  
In the remote 
environment, energy is expensive, so power minimization is another 
key concern.   
The DOM derives its internal power, including the PMT high
voltage, from the nominal $\pm$48V DC supplied by the cable.  Under normal operating conditions, power consumption is about 3.5 Watts/DOM.
All communication with a pair 
of In-Ice DOMs occurs over a single twisted copper pair; this includes power distribution,
bidirectional data transmission, and timing calibration signals.  

Most of the electronics reside on the Main Board (MB), which holds the
analog front-end and two digitizer systems.  The fast
digitizer system uses a 128-sample
switched-capacitor array, implemented in a custom Analog Transient
Waveform Digitizer (ATWD) chip, which can run between 200 and 700
mega-samples per second (MSPS).  
The ATWD sampling frequency is controlled with a 
digital-to-analog converter (DAC).  
A calibration procedure measures the ATWD sampling rate by
capturing the MB oscillator 20MHz signal and counting the ATWD bins
between oscillation cycles at several sampling frequencies, resulting
in a linear relation between sampling frequency and DAC setting.
Two ATWDs are used in a ping-pong fashion
to minimize dead-time.  The slow digitizer system uses a commercial flash ADC,
 operating at 40 MSPS digitizer, and allows a capture window of 6.4$\mu s$.

The digital functions on the MB are performed by a
400K-gate Field Programmable Gate Array (FPGA) containing a 32-bit
$ARM^{TM}$ CPU, 8 MBytes of flash storage, and 32
MBytes of random access memory.  
Aside from a small non-volatile
boot-up program, the operating parameters of the DOM, FPGA code, and
ARM software are all remotely reconfigurable.  
Timing on the Main Board is controlled
by a 20 MHz quartz oscillator, which is doubled to 40 MHz.

Within a DOM, data acquisition is initiated when the PMT signal
exceeds a programmable threshold, typically 0.3 
photo-electrons (PE) for DOMs in the ice.
This trigger is given a coarse time stamp by the 40 MHz local clock. 
The trigger initiates acquisition 
of data by the ATWD and capture of data from
the 40 MSPS ADC, 10 bit parallel output pipeline ADC.  
The ATWD collects 128 samples of 10-bit data, and
the commercial ADC collects 256 samples of 10-bit data.
To capture the entire waveform, before and after the trigger was issued,  the signal to the ATWD is
delayed 75 ns, using a 11.2 m long strip-line on a separate, dedicated
circuit board.  
The delayed signal is split among three (of 4) input
channels of each one of the ATWDs with gains differing by successive
factors of 8.  In this manner the digitizers cover the entire dynamic range of
the PMT, which is linear up to currents of 400 PE/15 ns.
The fourth ATWD channel is used for calibration and monitoring.  

The ATWD
sampling speed is variable and is currently set at 3.3 ns/sample, 
allowing acquisiton of 422 ns long waveforms. 
The precise timing of a signal is determined from the waveform
referenced to the coarse time stamp.
This is supplemented by timing information from the 40
MSPS ADC, which continually samples the amplified and shaped output of the PMT.

The DOMs can operate in one of several local coincidence modes, to reduce the noise-trigger-related 
data traffic to the surface. String 21 currently operates in a nearest-neighbor local coincidence 
mode, in which a local coincidence pulse is transmitted from a DOM to its immediate neighbors, above 
and below, whenever its discriminator fires. The DOM transmits its data to the surface only when it 
receives a local coincidence pulse within 800 ns of the trigger, signaling that at least one of its 
neighbors also had atrigger.

The flasher-board is an optical beacon integrated into each DOM. It contains
12 gallium nitride LEDs pointing radially outward from the DOM, 6 of which 
horizontally and the other 6 pointing upwards at an angle of 48$^{\circ}$.  Each LED is
capable of producing $1 \times 10^{7}$ to $1 \times 10^{10}$ photons per
pulse with peak wavelength in the range of 400 nm - 420 nm at a repetition
rate up to 610 Hz.
Special
operating modes allow an arbitrary ensemble of DOMs to be scheduled to emit
optical beacon signals into the ice for (1) calibration of  local coincidence,
timing and geometry;
(2) inter-string timing and geometry calibration; (3) verification of optical properties of the ice;
(4) linearity calibration of surrounding DOMs; (5) high energy cascade
calibration; etc.

A more detailed description of 
the operation of the DOM and its components is given
elsewhere~\cite{Stokstad,RealTime}.
\subsection{Production and testing}
\label{section:ProductionAndTesting}
IceCube DOMs are assembled at three different facilities worldwide.
Each site produces complete DOMs from
the incoming materials and tests them.  The total production rate from
all three assembly sites is in excess of 50 DOMs per week.

Because it is impossible to replace a bad unit or access it for
repairs, careful testing is essential.  The DOMs are tested in three
stages.  Before final assembly the separate subsystem components are
tested.  Fully assembled DOMs must then pass a series of performance
tests in special dark freezer laboratories where up to 120 DOMs can be tested simultaneously. 
The tests mimic the temperatures, the data acquisition procedures and
the optical signals that the DOMs will experience in the field.  
The DOMs are tested at four different temperature settings: 
room temperature, -45 $^{\circ}$C and -20$^{\circ}$C, which correspond to the temperatures of the 
highest and lowest DOM in the string, and -55$^{\circ}$C, which corresponds to the surface 
temperature in winter of the IceTop DOMs. 
The freezing and warming cycle is repeated twice. The entire test cycle takes 3 weeks.
DOMs that pass all the testing are 
shipped to the South Pole, where they are retested inside their shipping 
crates before deployment in the ice.

In addition to certifying DOMs as good for deployment, the testing 
measures critical performance characteristics for all DOMs in a controlled
laboratory setting~\cite{DOMpaper}.  In particular, short laser pulses create
single PE events which reveal a time resolution between 2 and 2.5 ns,
with 8\% of hits in a tail extending to 60 ns~\cite{RealTime}.
The late pulsing is a known PMT effect ~\cite{PMT}, and is due to electrons that were elastically scattered on the dynode, and have longer flight time in the PMT. This delay is less than expected for late arriving photons due to ice scattering (see Fig.  \ref{nures}).

\subsubsection{Data acquisition software}
The DAQ testing software consists of
Java-based code interacting with software and
firmware running on the DOMs. The Java software runs simultaneously on
multiple host computers, which are tied together using Java's Remote Method
Invocation.
DOM configuration is done using an XML file that defines global run and individual DOM parameters.
 There are about 50 settable parameters
per DOM. Examples of these parameters include the
length of the run, high voltage setting, ATWD sampling rate,
the number of samples to read out for each ATWD channel, 
the DOM trigger mode operation, the discriminator set
point, flasher-board setting, etc. 
Standard settings for each DOM are stored in a
master database.  
Each DOM has a unique electronic read-back
identifier, which is mapped both to a string number and to a position,
as well as to a simple-to-remember name.  

Perl-based experiment control scripts run on a special designated computer called a string processor. The string processor communicates with the DOMs using a designated DOMhub computer that can configure each of the DOMs according to XML configuration files. Three data streams are open for each DOM for hit records, monitor records and time calibration records and the data collected for each DOM is written to three corresponding files. The GPS event time is added to each record.

The same software was also used for DAQ at the South Pole during the first year of operation
discussed in this paper. After a South Pole run was completed, 
the data were filtered in a software.  The In-Ice filter required 8 DOMs that launched within
two microseconds.
This reduced the data volume from about 30 GBytes to less than 4 GBytes per day.
The reduced data were transmitted to the northern 
hemisphere for further processing.

\begin{figure}[htb]
\begin{center}
\begin{tabular}{ccc}
\mbox{\includegraphics[width=8.5cm]{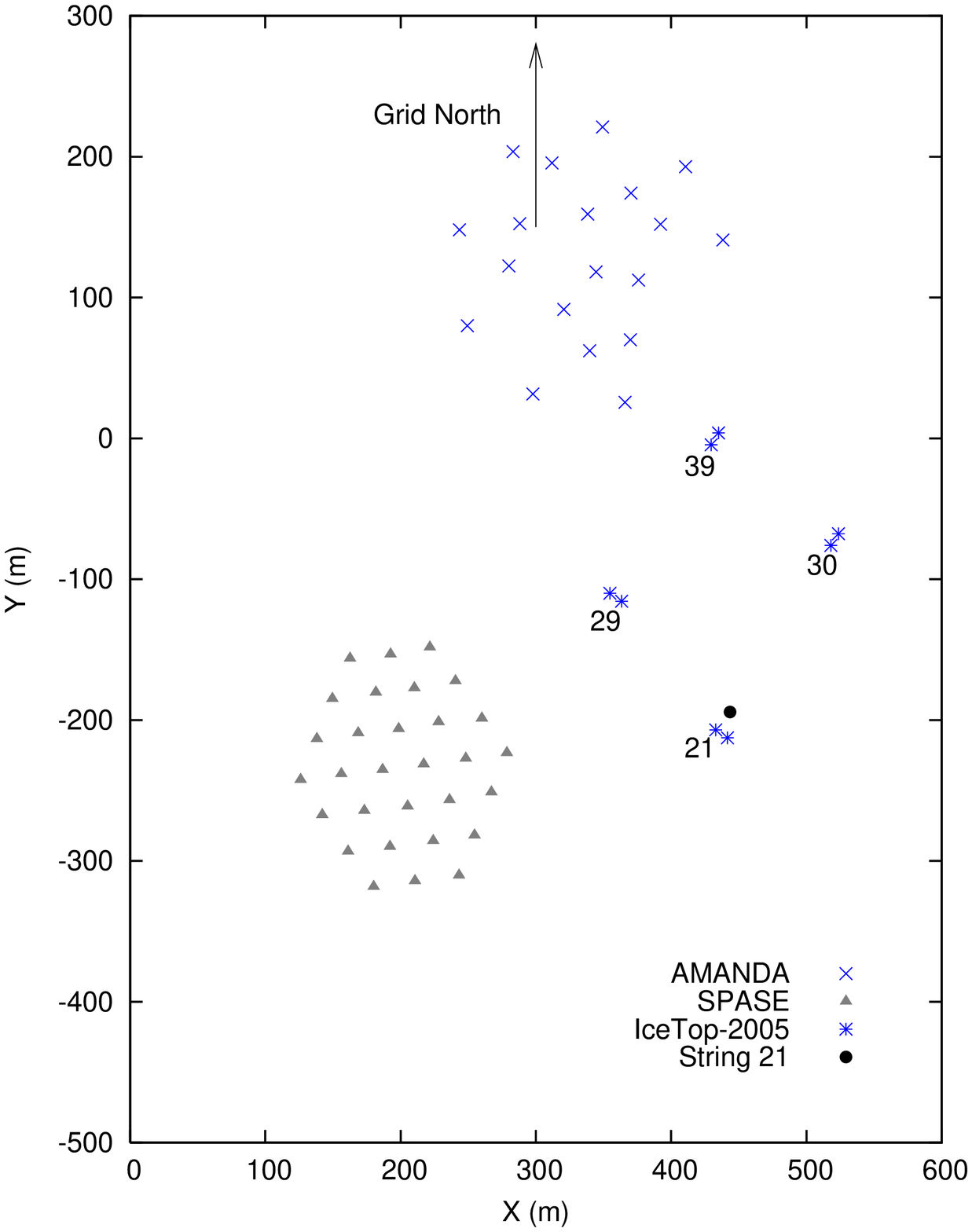}} & \ & \mbox{\includegraphics[width=6.5cm]{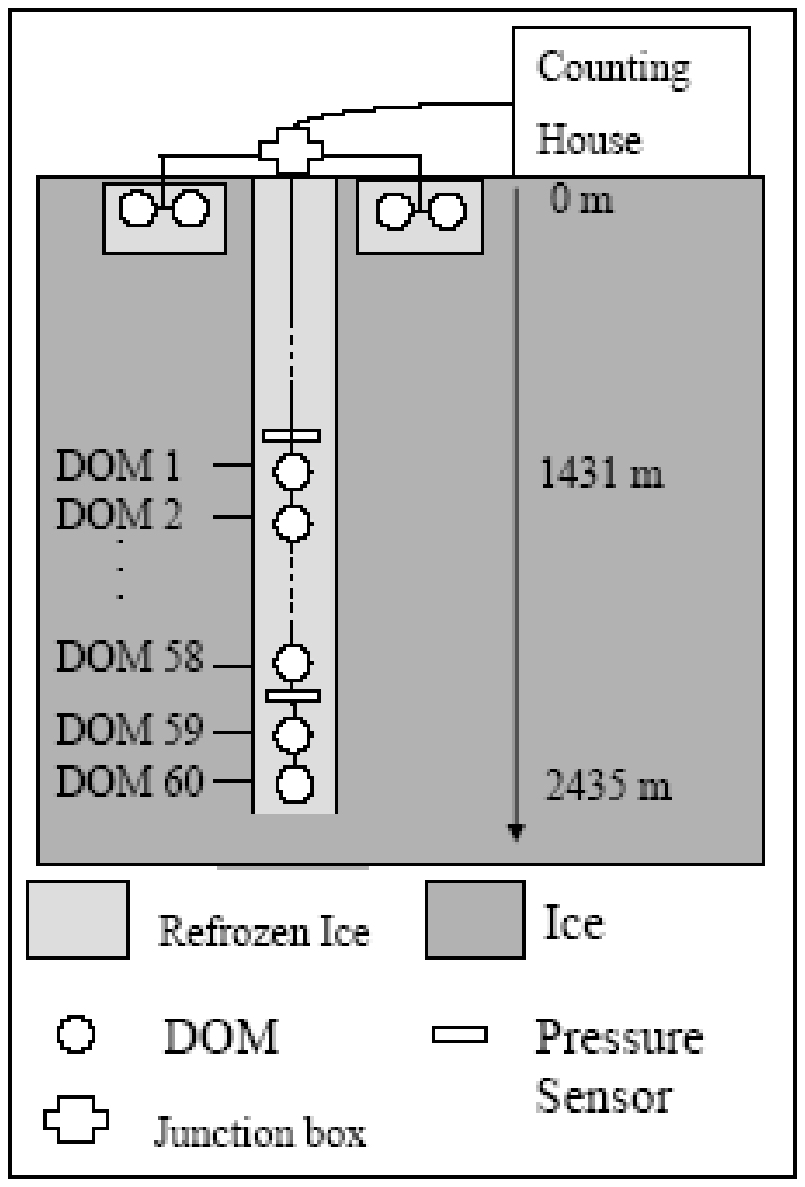}} \\
\parbox{.45\textwidth}{\caption[]{\label{map05+A} 
IceCube in 2005 - Surface map of IceTop and String-21 with AMANDA and SPASE.
}} & \ & \parbox{.45\textwidth}{\caption[]{\label{hole} 
Schematic diagram of String-21 and two associated tanks. }} \\
\end{tabular}
\end{center}
\end{figure}

\subsection{Configuration in 2005}
\label{section:configuration}
In December and January 2004/2005 the first IceCube string was deployed
along with four IceTop stations (eight tanks containing a total of 16
DOMs).  This string is at location 21 of 80 on the
IceCube plan and is therefore referred to as ``String-21".
Fig.~\ref{map05+A} shows a surface map of the 2005 configuration, 
including the 19 AMANDA strings~\cite{AMANDA} and the 30 stations of 
the South Pole Air-Shower Experiment (SPASE) ~\cite{SPASE}.  The four 
IceTop stations enclose an area of approximately $1.7\times 
10^4$~m$^2$.
\subsubsection{Drilling and deployment}
\label{section:DrillingAndDeployment}
Strings of modules are deployed into holes drilled with hot water.
The most critical and demanding component of the IceCube deployment
process is the forced hot water drill system.  
The plant for producing the drill water comprises a total of 15 buildings that host the hot water heaters, generators, pumps and storage tanks. 
These buildings are set up in a temporary equipment camp during each season.

A supply hose from the equipment camp to the drill site
connects to the hose winch, which can support a single 2.5 km long drill hose.  
The nozzle of
the hot water drill is connected to the end of the drill hose and
lowered into the hole at the rate ice is melted by the hot water.  A return
hose carries cool water from the hole back to the heating system.
Cables for both drilling and deployment are fed through a hole in the
roof of a tower operating structure, which is a building located
over the hole.  The tower is heated and contains a control room for
drilling and deployment operations.  A water-filled hole is prepared to a depth of 2.5 km 
with a minimum diameter of 60 cm.  This is wide enough to insure a 45 cm diameter
cylinder of water remains when the string of DOMs is later deployed.
The drill hose is
removed and rewound on the hose winch, and the tower is then
available for deployment.

After building up the equipment site and assembling the tower and drill system for
the first time, the hole was ready for string deployment on
January 27, 2005.  The deployment of String-21 took a total 
of 18 hours.  The following day, the string was connected 
to the junction box, and commissioning began.

\begin{figure}[htb]
\begin{center}
\includegraphics[angle=270,width=12cm]{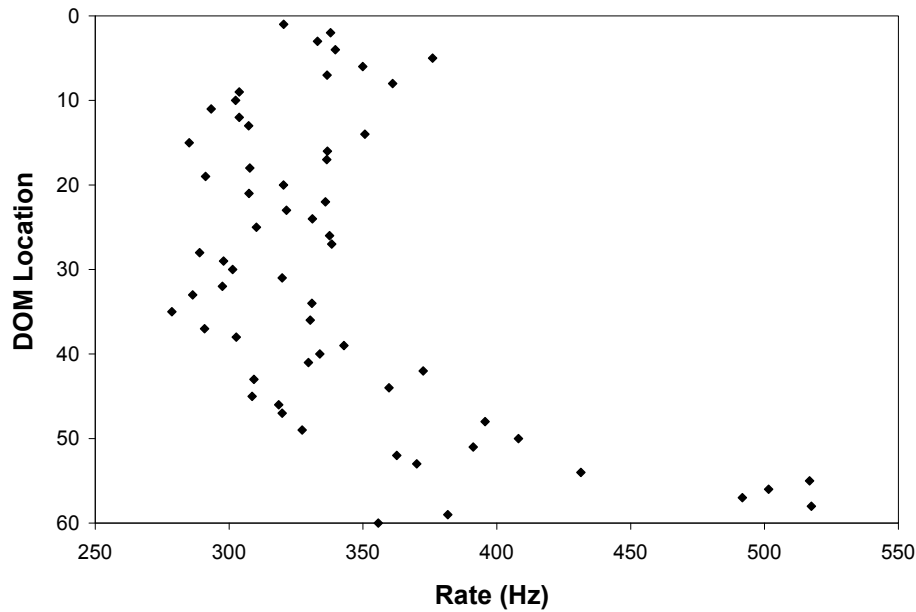}
\end{center}
\caption{Noise Rates when after-pulsing is suppressed.  DOMs are numbered
in sequence from 1 (highest) to 60 (lowest).}
\label{String21-with-51us-deadtime}
\end{figure}

Soon after deployment the PMT rates rose dramatically
because of triboluminescence caused by stresses in the ice forming near the DOMs as the water in the borehole refroze.
The time for the water in the hole to refreeze varied 
with depth, from 5 days at 1.5 km to more than two weeks at 2.5 km.
After the refreeze was complete, 
DOM rates decreased to a typical noise level of 650 Hz. When a dead-time 
interval is enforced to remove pulsing
correlated in time within the same PMT, the noise is reduced further. 
This is
shown in Fig. ~\ref{String21-with-51us-deadtime}. 
Such a low noise rate is favorable for searches for low energy neutrinos
from stellar collapse, the detection of which would follow from an observation of a simultaneous increase in the overall counting rate~\cite{SupernovaSearch}.
\subsubsection{The In-Ice Configuration}
\label{section:InIceConfig}
String-21 consists of a 2540 meter cable, 60 DOMs,
pressure sensors, thermistors, and a dust logger to measure the optical 
properties of ice throughout the deployment~\cite{dustlogger}, as shown 
schematically in Fig.~\ref{hole}.  
The 60 DOMs are spaced 17 meters 
apart on the deepest 1000 meters of the string.

The actual depth of the deepest DOM on the
 string (2435 m) was determined with data from the two high-resolution pressure
sensors, spaced 1000 m apart, together with a laser survey of the distance from the surface to the water level in the borehole. 
The pressure reading from the deeper pressure sensor, located just above the next-to-deepest DOM,  
provided the absolute depth of the string, 
which is defined as the depth of the deepest DOM. The pressure was
corrected for the compressibility of the water in the hole, which was
measured with data from both pressure sensors by requiring that the
difference in depth reading be independent of depth and in agreement with
the nominal spacing of the sensors (except for a small difference due
to cable stretching). The coordinates of the DOMs higher up the string were then determined by
adding the vertical spacing between modules, which were measured during deployment with a laser ranger.
Depths of individual DOMs are determined to an
accuracy of 50 cm.
\subsubsection{IceTop Tanks } 
\label{section:IceTopAssembly}
About 25 m from the top of each string there will be an IceTop station.
There are two tanks per station separated by 10 m. 
The 80 IceTop stations will form an air-shower 
array with a nominal grid spacing matching the 125 meter string 
spacing.  The spacing between tanks at a station is chosen to maximize 
the probability that single-station hits 
are caused by small cosmic-ray showers 
that contain only one muon capable of penetrating to the deep detector.  
(A single-station event is defined as one in which there is a
 coincidence between the two tanks at one station with no hits 
in adjacent stations.) 
Such cosmic-ray induced muons are the main background for a 
neutrino telescope. 

Tanks are deployed in shallow trenches so that,
after backfilling, the tank tops are initially at the surface (to
minimize buildup of drifting snow).  At the bottom of the trench
between each pair of tanks is a surface junction box to which the
associated IceTop DOMs and hole cable are connected before
backfilling.  Tanks are cylindrical with 1.8~m diameter.  The depth of
ice in the tank is 90~cm.
Standard survey techniques are used to determine the tanks position to within 5 cm.

Each tank is viewed by two downward facing DOMs with their lower hemispheres
embedded in the ice.  There are two DOMs per tank for flexibility and 
redundancy.  Operating the two PMTs in a tank at different gains allows 
an increase in dynamic range beyond what is available for a single gain
setting.  Alternatively, the DOMs may be adjusted to the same gain to give 
a more uniform tank response to incident particles.  In the data 
described here, one DOM was operated at a gain of approximately 
$5\times 10^6$ and the other at approximately $5\times 10^5$.  With 
this setting, the signals generated by single through-going muons lie 
within the range of linear response for both the high-gain and 
low-gain DOMs.
\subsubsection{Cabling}
\label{section:Cabling}

The cable system was designed to maintain a Faraday shield around the 
system and to provide optimum performance subject to constraints 
imposed by the conditions of delivery and deployment.

The cable system consists of the following main components:
Surface to DOM Cable Assembly, Surface Junction Box Assembly, IceTop Cables, 
Surface Cable Assembly and Counting House Cable Assemblies.
The cables meet at the Surface Junction Box, which is made of stainless
steel and has a removable lid.                                          
The Surface to DOM Cable Assembly is the largest and longest cable in 
the system. It provides power and communications to the DOMs and all other In-Ice sensors from the Surface Junction Box. To reduce the weight and cost 
of the cabling, each pair of adjacent DOMs communicates with the surface 
over a single twisted pair.  Two twisted pairs are combined to form 
a ``quad''.
The cable is 2540 meters in length, contains 18 quads, four bundles of 
3 twisted pairs, and two 25 kN strength members.  One of the strength 
members runs down the center of the cable and the other is woven into 
the outer sheath.

There are 30 breakouts, each providing two connection points for DOMs. 
A breakout is made by slicing open the outer sheaths of the cable, 
cutting quads and service wires, attaching the wires to connectors, 
and re-sealing the cable.  The connectors used for DOM connections are 
waterproof to 10,000 psi (680 atm).  For the connections on the Surface Junction Box end of the cable, the 
quads and twisted pair bundles are attached to military specifications, round, metal 
shell connectors.

Since the DOMs all communicate independently, cross-talk is a
potential source of errors in communication and timing.  Cross-talk
suppression requires careful mechanical assembly of the cable, as
mechanical asymmetries or imperfections in the twisted quad
configuration introduce cross-talk.  Care is also required during
deployment to avoid distorting the cable elements.  The requirements
for the cable are that near-end and far-end cross-talk be suppressed
by more than 50 dB and 30 dB, respectively.
\subsection{Data Acquisition Hardware}
\label{section:ControlAndDAQ}
The South Pole host machines comprise DOMHubs, which are standard 
industrial Single Board Computers, and a string 
processor, which is a commercially available server computer. Each 
DOMHub is customized with +48 Volt and -48 Volt switching regulated AC-DC single output power 
supplies, to supply 96 Volts to the DOMs.  Wire pairs servicing DOMs are connected  
to custom PCI cards inside the DOMHubs called DOR (DOm Readout) 
cards. The DOR cards provide power, communications and time calibration signals to the 
DOMs. 
Each DOR card can be connected to two power and communication 
wire pairs.  Each wire pair is connected to two adjacent DOMs 
on a string (one terminated 
and one unterminated to optimize communications).  IceTop
DOMs are served by a single wire pair.  
A DOMHub with a 
full complement of eight DOR cards can read out up to 32 DOMs
In-Ice or 16 IceTop DOMs. 
The DOR cards used in 2005 have been
replaced with a new version capable of handling 2 quads (8 DOMs), such that each DOMHub serves one complete string.

Each DOMHub is equipped with a custom PCI card called a
DSB (DOR Service Board) card. The DSB card distributes the GPS time
string to each of the DOR cards in a DOMHub.  These signals are used
to maintain timing across the array. 
\section{Calibration of gains and timing}
\label{section:Calibration}
Several steps are necessary to transform the DOM readouts to waveforms
in physical units of charge and time.
Section \ref{section:ATWDCalibration} describes the DOM specific
calibration of the ATWD sampling frequency and voltage scale and the
measurement of the different gains of the three ATWD channels.  Section
\ref{section:CalibrationPMTGain} will describe the PMT gain
calculation. The entire procedure is carried out by DOM-Cal, a
software package executed on the embedded DOM CPU.  In section
\ref{section:TimingCalibration} the time calibration between the
different DOMs and an absolute reference time are described.  
Section \ref{section:flashers} describes calibration with flashers, and 
in section
\ref{section:WaveformReco} we discuss the calibrated waveforms measured by a DOM.
\subsection{The ATWD and Amplifier Calibration}
\label{section:ATWDCalibration}
The DOM ATWD bin readout scales linearly with input signal voltage;
however, each ATWD bin has a unique slope (V/ADC-counts) and intercept.
DOM-Cal measures this response by varying ATWD bias voltage and 
applying a linear fit to the resulting amplitudes
for each bin (0-127) of each signal channel (0-2) for both ATWDs.
Each ATWD bin can be digitized to a depth
of 10 bits.

The DOM front-end amplifier gains for each channel are 16x, 2x, 
and 0.25x, providing complete dynamic range for signals generated by the PMT. 
The amplifier gains vary due to component tolerances and must be determined precisely for PMT gain calibration.  
Amplifier calibration is a two step process:  The absolute gain
of the high-gain channel is determined first, then the lower-gain
channels are calibrated relative to the high-gain channel.
The high-gain channel is calibrated by injecting artificial Single Photo Electron (SPE) like
pulses into the DOM PMT input and comparing peak amplitude from the
ATWD to the true peak amplitude of the pulse.  
The amplifier gain is given by the mean peak ratio for a large sample of
pulses.  ATWD sampling speed is maximized during the procedure
to minimize waveform integration error.
\begin{figure}[htb]
\begin{center}
\begin{tabular}{ccc}
\mbox{\includegraphics[angle=270,width=.45\textwidth]{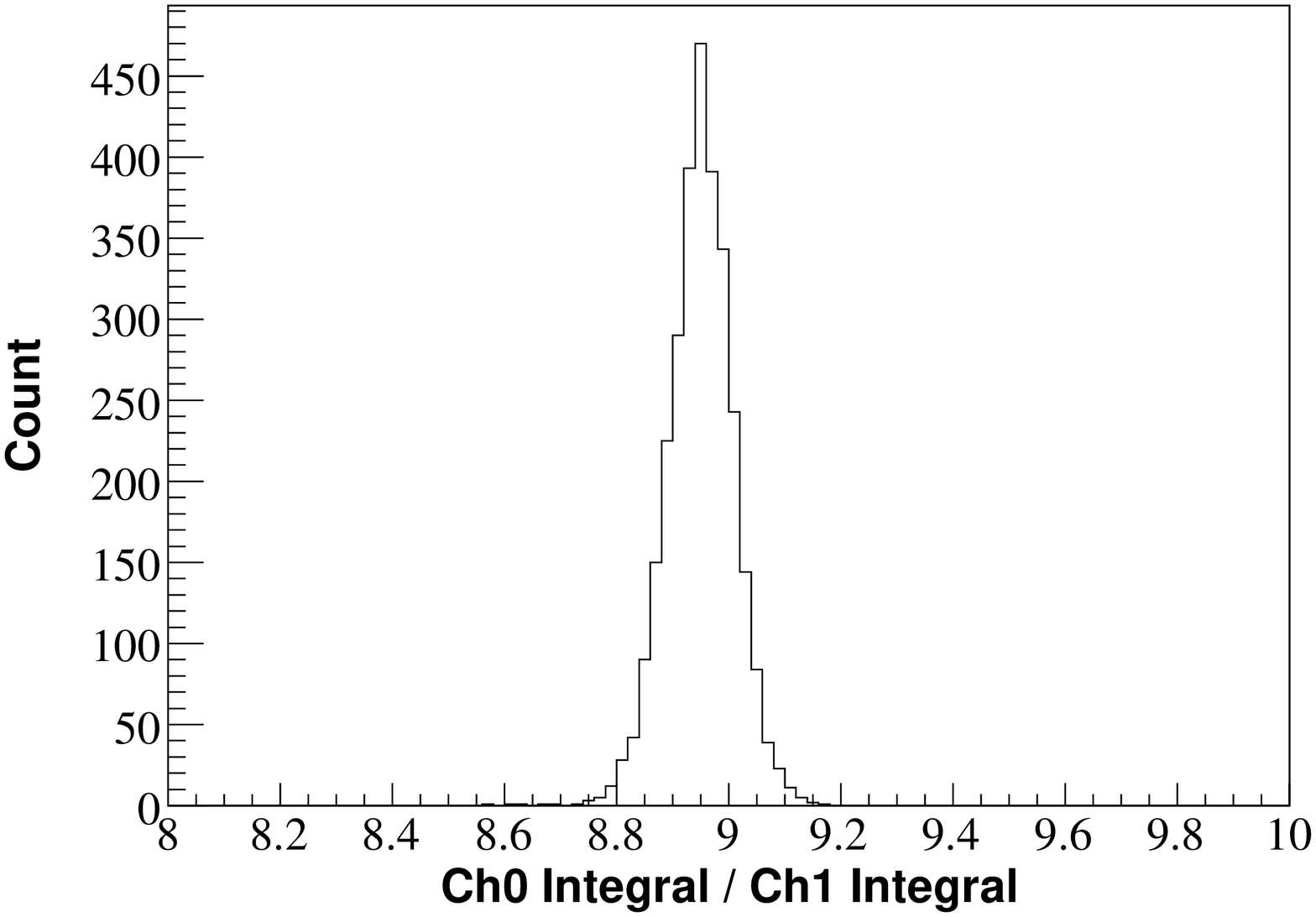}} & \ & \mbox{\includegraphics[angle=270,width=.45\textwidth]{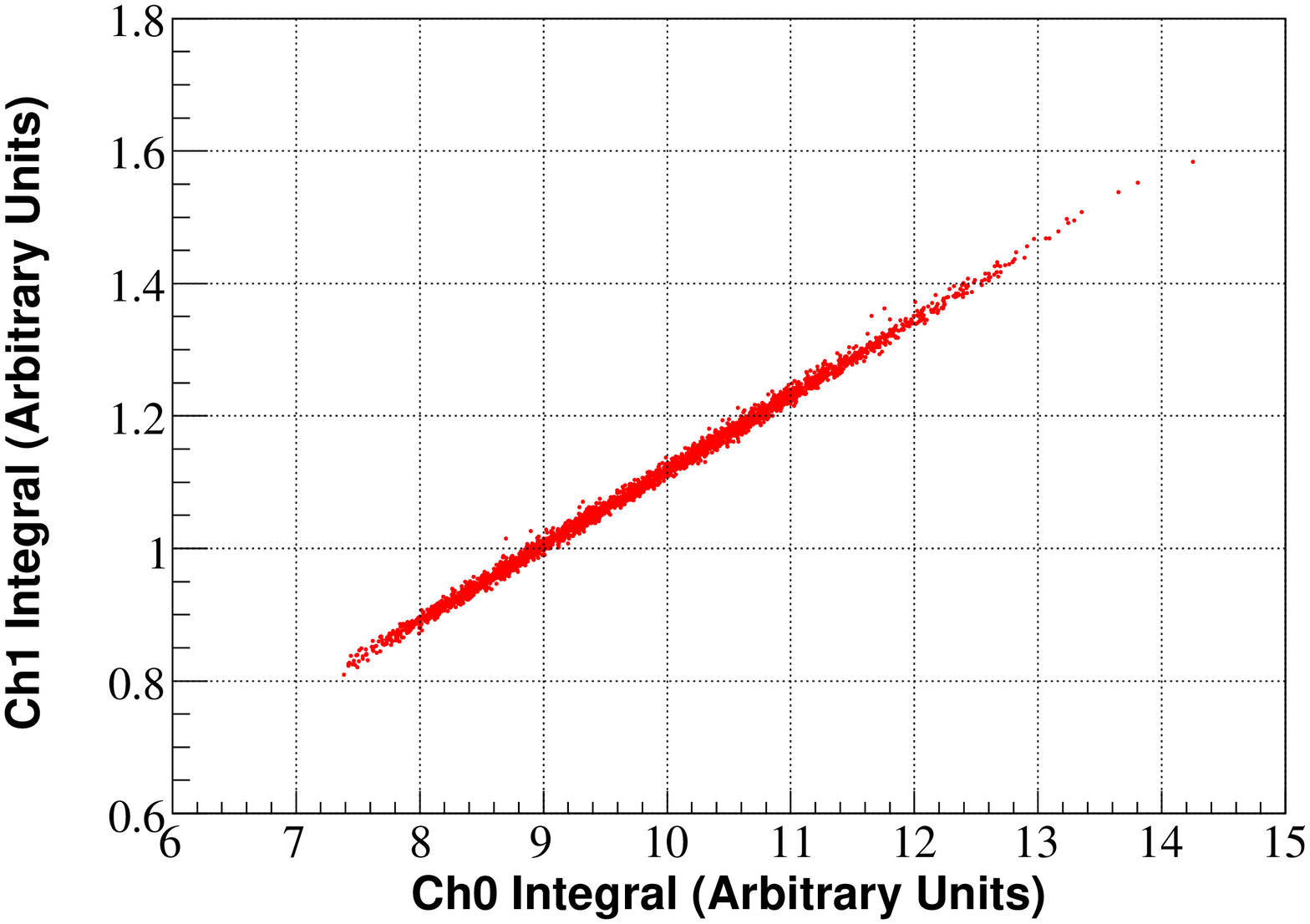}} \\
\parbox{.45\textwidth}{\caption[]{\label{ch0ch1} Ratio of ATWD channel 0 (high-gain) waveform area to ATWD channel 1 (moderate-gain) waveform
area for 3000 LED pulses.}} & \ & \parbox{.45\textwidth}{\caption[]{\label{ch0ch1line} ATWD channel 1 waveform area vs. ATWD channel 0 waveform area.}} \\
\end{tabular}
\end{center}
\end{figure}

To calibrate the lower-gain channels, DOM-Cal uses PMT signals from
light pulses generated by a LED on the DOM MB.  ATWD data is
acquired and calibrated both for the channel with the unknown gain and
the previous channel, whose gain was already calibrated.  The gain of
the next channel is the gain of the previous channel times the ratio
of the pulse integrals of the two channels.
LED pulse ranges are chosen to avoid ATWD saturation in the higher-gain
channel, yet still provide significant amplitude in the lower-gain channel to
minimize integration error.
Fig. \ref{ch0ch1} and Fig. \ref{ch0ch1line} show the relation
between the gains of two ATWD channels.
\subsection{Calibration of the PMT gain}
\label{section:CalibrationPMTGain}
Both DOM acceptance testing and detector operation require
the DOMs to operate at a specified gain.  Each PMT has a unique
gain response to anode-photocathode voltage; therefore, it is
necessary to characterize this response and calculate the 
voltage yielding a specified gain for each DOM.  

\begin{figure}[htb]
\begin{center}
\begin{tabular}{ccc}
\mbox{\includegraphics[width=.45\textwidth]{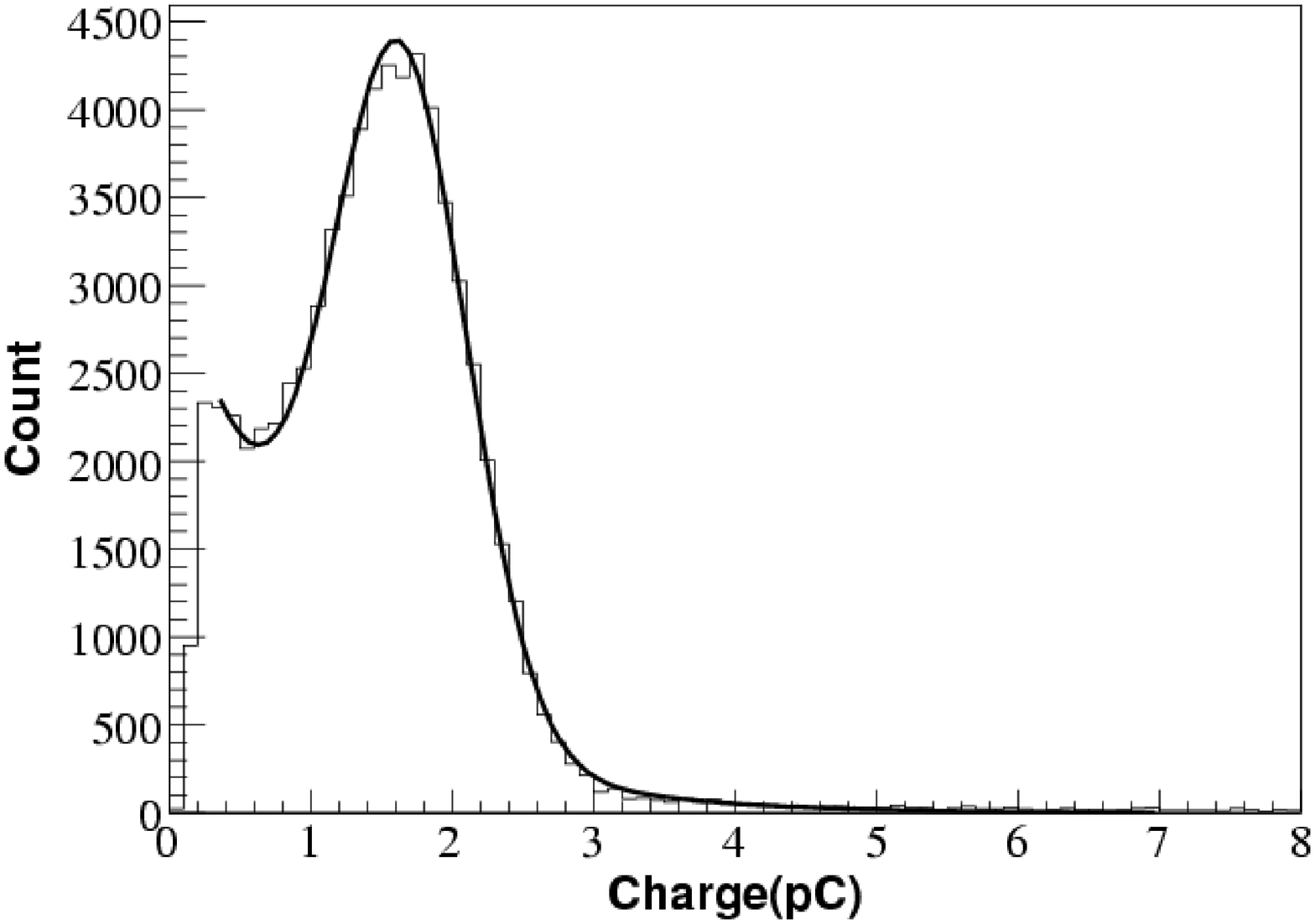}} & \ & \mbox{\includegraphics[angle=360, width=.45\textwidth]{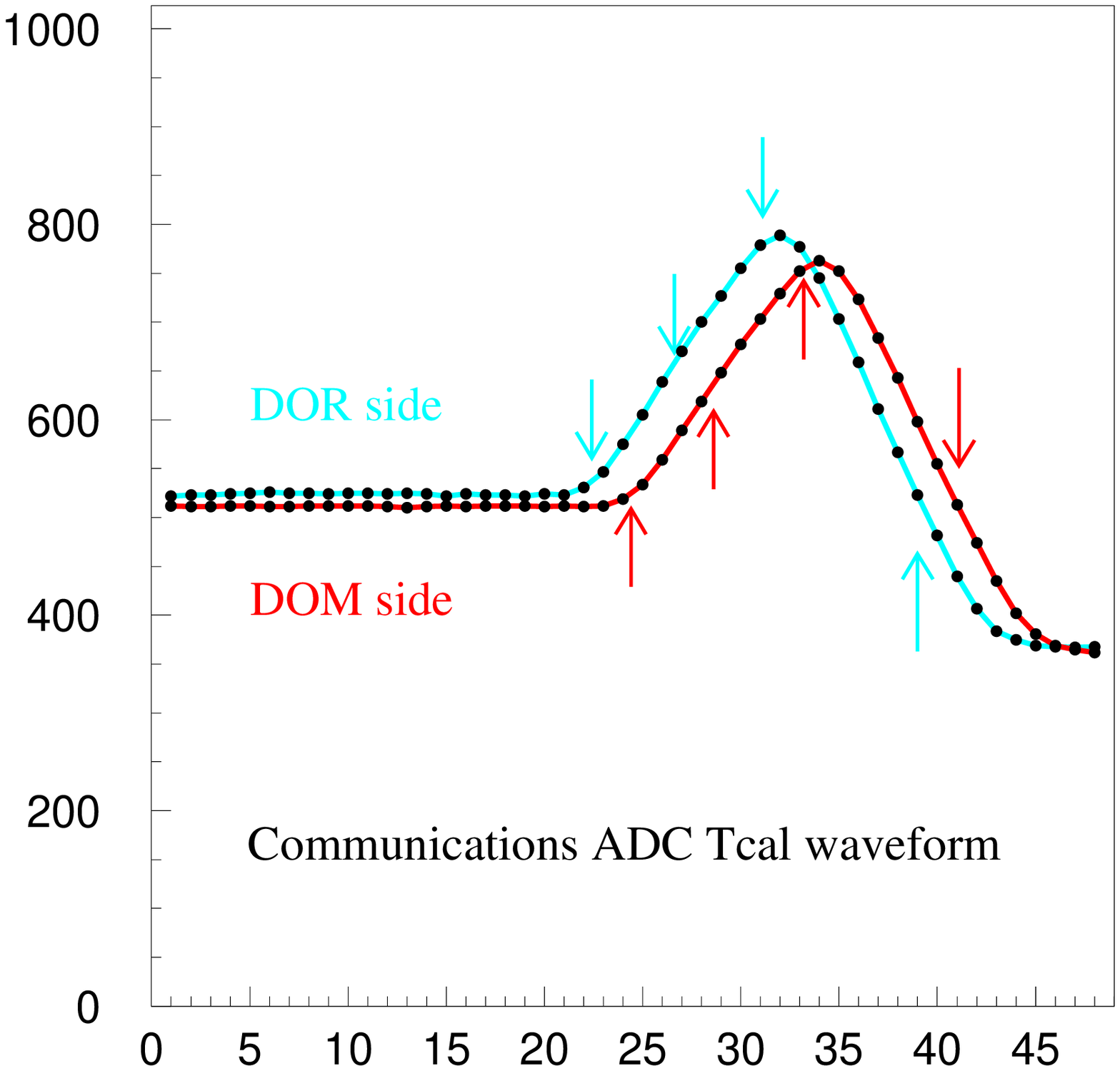}} \\
\parbox{.45\textwidth}{\caption[]{\label{disc} A typical DOM discriminator spectrum at $10^7$ gain.  The spectrum is dominated by the Gaussian distribution of single photoelectrons.}}& \ & \parbox{.45\textwidth}{\caption[]{\label{at} Time calibration waveform as measured by the DOM and at the surface after propagation through the cable. Features used in the time calibration are pointed out.}}\\
\end{tabular}
\end{center}
\end{figure}

The uncorrelated noise in the optical module results from thermal 
background of the photocathode, which is
significantly reduced at the cold temperatures, 
and by radioactive decay of isotopes contaminating the glass pressure sphere. 
The electrons from the beta decays produce
photons by Cherenkov radiation and by scintillation. 
This results in a significant rate of SPE signals.
The spectrum of PMT signals is shown in Fig. \ref{disc}.

At each high voltage setting, several thousand PMT waveforms are acquired with a special low discriminator threshold. Each waveform is integrated around the peak. The total charge is obtained by dividing the integrated value by the load impedance and the ATWD sampling frequency.
The charge spectrum is fitted to a Gaussian + exponential function. The mean SPE charge is proportional to the mean of the Gaussian component. The gain is the SPE charge divided by the electron charge.
This procedure is  repeated for high voltage settings between
1200V and 1900V in 100V increments. A linear fit of $log_{10}(gain)$ to $log_{10}(V)$ is performed.
This fit is later used to compute the PMT voltage yielding a desired gain.
The gains of the PMTs on String-21 were set at $10^7$.

\subsection{Timing calibration}
\label{section:TimingCalibration}
The time calibration of the DOM is based on a procedure called 
Reciprocal Active Pulsing~\cite{Stokstad}, in which the phase and frequency of each DOM's 
free running local oscillator (20 MHz) are determined relative to a master 
GPS-controlled oscillator at the surface by transmitting a fast bipolar 
pulse at known intervals from the DOR card to the DOM.  
After receiving this pulse the DOM "reciprocates" 
and sends an identical fast bipolar pulse after a known delay interval to the DOR card.  
The time calibration waveforms are produced and digitized by the same DACs and ADCs 
(operating at 20 MSPS) used for digital communication. 
The 2.5 km long cable slows the rise time of the original fast calibration 
pulse from a few nanoseconds at the source to a microsecond or more at the 
receiving end. Fig. \ref{at} shows the received time-calibration waveforms. 

The accuracy of this calibration method arises from the reciprocity or symmetry of the 
pulsing system, in which the pulses generated at each end have the same shape and the 
dispersed, attenuated pulses received at each end have the same shape.  In this limit, 
the one-way transit time is equal to one-half the round-trip time minus the known delay, 
regardless of which feature of the waveform (e.g., leading edge or crossover) is taken 
as the fiducial mark.  Reciprocity is achieved because the calibration signals follow 
the same electronic path through the same components for both transmission and reception.  
Reciprocity is also verifiable, because the calibration waveforms are digitized in both 
the DOM and DOR; these waveforms can be compared to determine any differences in shape 
that might introduce systematic errors.

\begin{figure}[!h]\begin{center}
\begin{tabular}{ccc}
\mbox{\epsfig{file=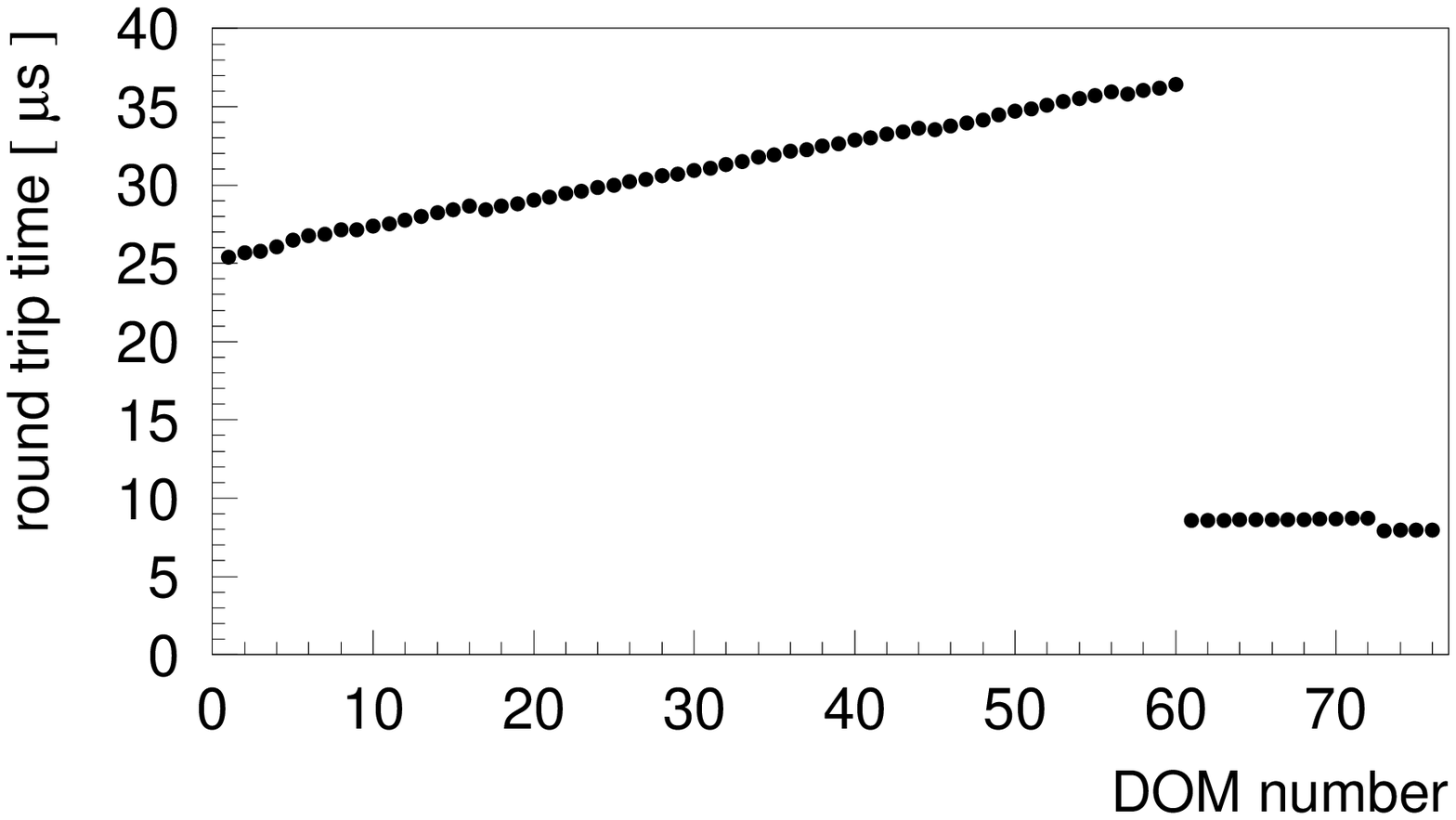,width=.45\textwidth}} & \ & \mbox{\epsfig{file=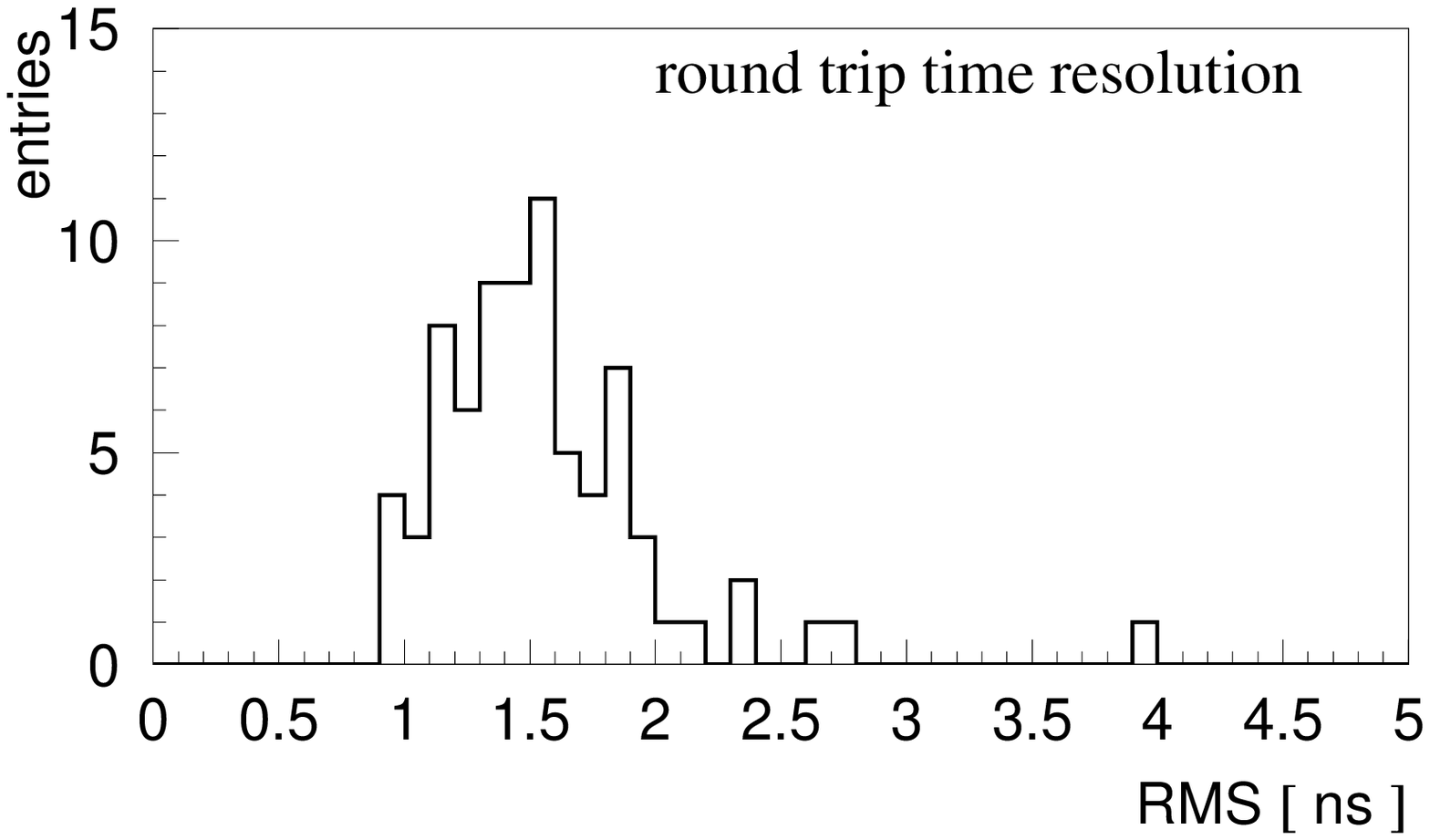,width=.44\textwidth}} \\
\parbox{.45\textwidth}{\caption[]{\label{a1} Round trip time of the time calibration 
pulse. }} & 
\ & \parbox{.44\textwidth}{\caption[]{\label{a2} The rms variation of the round trip time of the time calibration pulse for all 76 DOMs.}} \\
\end{tabular}
\end{center}\end{figure}

The round-trip travel times for different DOMs depend on the electrical length of the 
cable connecting the DOR and DOM. This is illustrated in Fig. \ref{a1}. The transit 
times increase for deeper sensors on the string (DOMs with numbers 1-60), and are 
similar to each other for the surface DOMs (shown as numbers 61-76).  
At present, calibrations are done automatically every three seconds, 
although longer intervals up to ten seconds may be used in the future. 
The variation in the round-trip travel time from one calibration to the next provides
the basic measurement of the precision of the time calibration procedure.  
The best resolution is obtained using crossover timing, for which the 
typical round-trip resolution is less than 2 ns RMS (Fig. \ref{a2}).  

An LED on the main board is used to measure the photo-electron transit time in the 
PMT and other delays in the photon signal path.  The statistical and systematic 
precision with which photon arrival times can be determined depends on 
additional factors, such as PMT performance and waveform analysis. 
This is measured in experiments using flasher signals and down-going muons to be 
described in subsequent sections.
\subsection{Time resolution studies using the DOM flashers}
\label{section:flashers}
\begin{figure}[!h]\begin{center}
\begin{tabular}{ccc}
\mbox{\epsfig{file=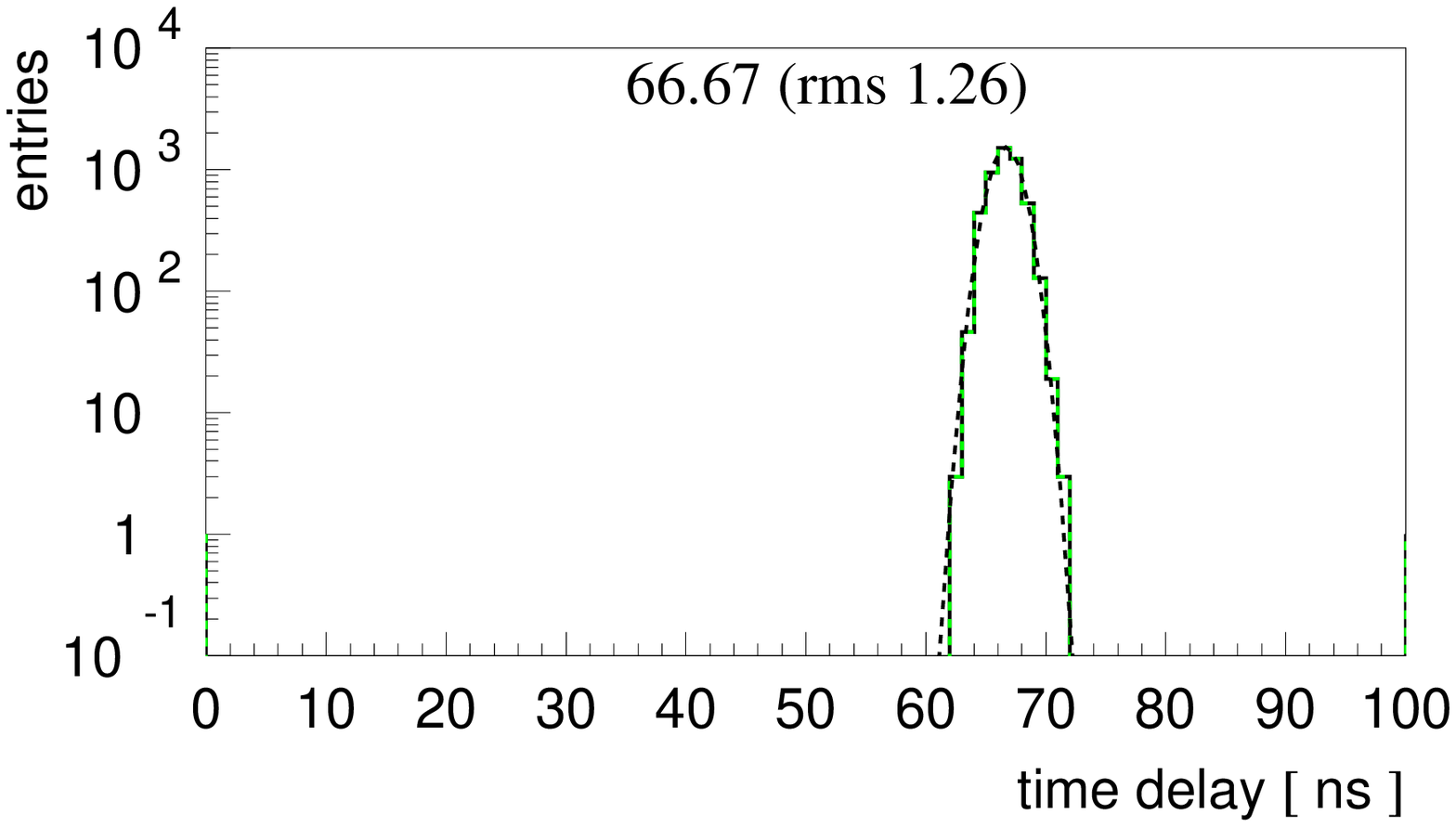,width=.45\textwidth}} & \ & \mbox{\epsfig{file=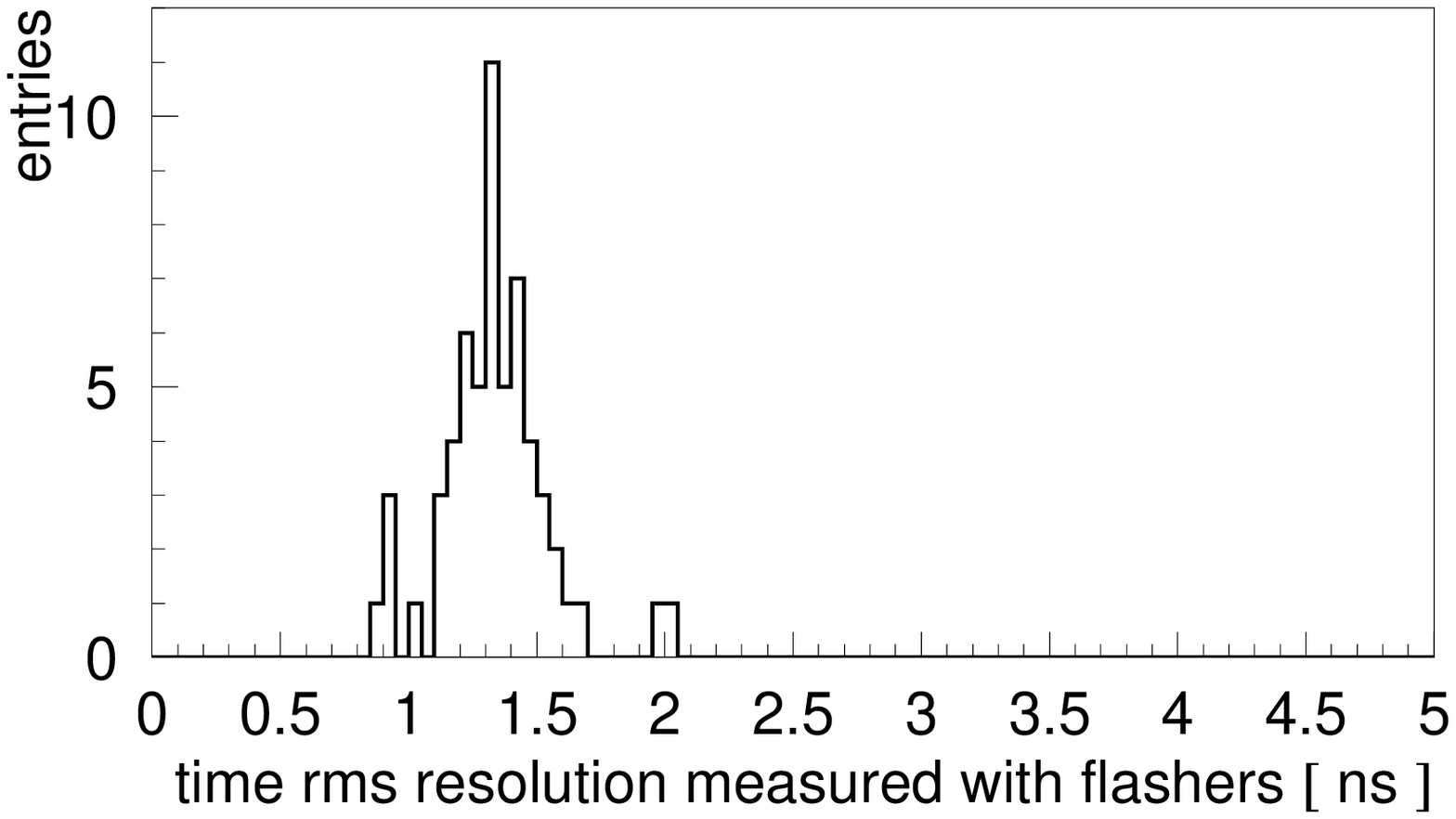,width=.44\textwidth}} \\
\parbox{.45\textwidth}{\caption[]{\label{ab} Photon arrival time delay at DOM 46 when DOM 47 is flashing in clear ice.
The mean and rms values are indicated.}} & \ & \parbox{.44\textwidth}{\caption[]{\label{ac} 
RMS variation of time delay measured with flashers for 58 DOM pairs on the IceCube string. }} \\
\end{tabular}
\end{center}\end{figure}

The LEDs on the flasher-boards were used to measure the photon transit (or delay) time 
for the reception of a large
light pulse at the closest DOM above the one flashing.
Fig. \ref{ab} shows the distribution of 
delay times for DOM 46 when DOM 47 was flashing. 
The mean time delay is given by the light travel time from the flasher to the
receiving DOM, together with a small electronics offset.
The RMS variation
of the time delay,  
shown in Fig. \ref{ac} for 58 DOM pairs, should reveal any imprecision or drifts 
associated with the clock calibration procedure (\S~\ref{section:TimingCalibration}).
There is also a small contribution from scattering in the ice; the RMS values
are smallest for the DOMs located in clearer ice. 
\subsection{Waveform reconstruction}
\label{section:WaveformReco}

\begin{wrapfigure}{r}{7.8cm}
\vspace{-0.2cm}
\epsfig{file=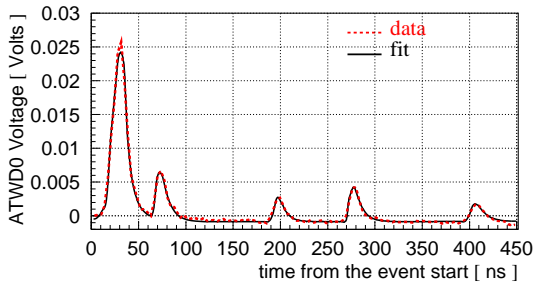,width=0.5\textwidth}
\vspace{-0.2cm}
\caption[]{\label{ay} Captured hit event waveform.}
\vspace{-0.4cm}
\end{wrapfigure}

A typical calibrated waveform captured by a String-21 DOM is shown in Fig. \ref{ay}. 
Different waveform features, such as time, amplitude, width and area of the primary and secondaries pulses can be used for reconstruction.
These waveforms can be described well by a 
decomposition procedure that yields an ensemble of single photon hit times. 
The procedure is based on iteratively fitting a waveform with a function 
that is a sum of a constant and a progressively larger number of terms each 
describing a single photo-electron response with allowed variations in amplitude 
and width. An alternative method applies a Bayesian unfolding 
algorithm to the waveform with single photo-electron response as a smearing function. 
Both methods give photon arrival times within 0.5 ns of each other.
\begin{figure}[htb] \includegraphics[width=14cm]{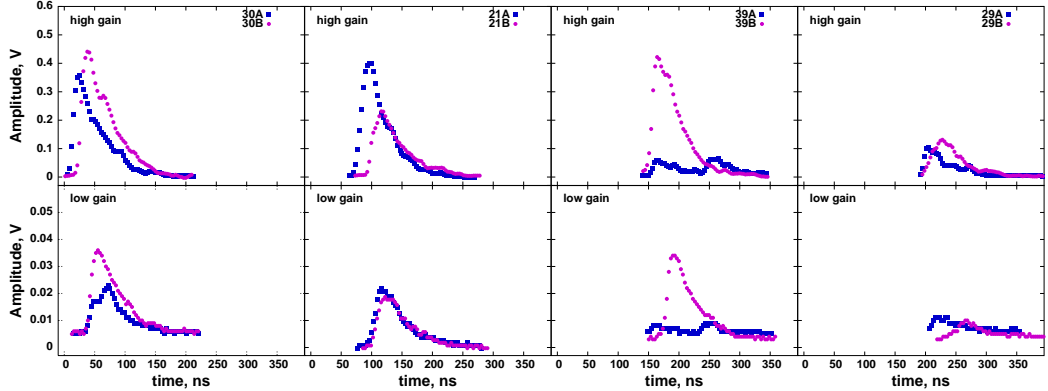}
\caption{Waveforms of an air-shower seen by all sixteen DOMs in the
4-station IceTop array.  The timing of
the signal in DOMs at different locations is used to determine the
shower arrival direction.}
\label{waveform}
\end{figure}

In contrast to the data generated by muons in the ice, the air
shower signals in the surface tanks are much bigger and require
different procedures.  Fig. ~\ref{waveform} shows the signals from an
air-shower seen by all sixteen DOMs in the array, as a function of
time relative to the first DOM hit.  Each panel shows traces from two different tanks
at the station.
The top row shows the waveforms seen by the high-gain
DOMs ($5\times 10^6$) in each tank while the bottom row shows the
low-gain ($5\times10^5$) data.  In the event shown here the 
DOMs have not yet been adjusted to give the same response for a
given number of photo-electrons.
The integrated waveforms
correspond to several hundreds of photoelectrons from the light produced 
by many particles
in the shower front hitting the tank in quick
succession. 
 
\section {Track reconstruction in IceCube}
Reconstruction of the tracks of muons, whether of
downward muons from cosmic-ray interactions in the atmosphere
or neutrino-induced muons, is essential for
a neutrino telescope.  Pions and kaons created by interactions
of cosmic-rays in the atmosphere give rise both to 
downward atmospheric muons and to upward-moving muons
from interactions of atmospheric neutrinos.  Comparison of these
two closely related and well-understood fluxes therefore offers an
important consistency check on detector simulation
and response.  The abundant atmospheric muons also
provide a useful beam for calibration of timing
and geometry.
\subsection{Muon track reconstruction}
\label{section:MuonReconstruction}
A maximum likelihood algorithm for one-string track reconstruction 
employing the full waveform unfolding method of 
Section \ref{section:WaveformReco} was used to reconstruct the In-Ice data. 
The likelihood function was built using an approximation to the photon arrival 
time distribution for multi-layered ice first introduced for the AMANDA reconstruction 
in \cite{wiebusch}, and refined in \cite{japaridze}. 
The approximation can be 
used for ice with variable properties (i.e., depth-dependent scattering 
and absorption) if scattering and absorption coefficients are averaged 
between the photon emission and reception points in ice~\cite{japaridze}.

\begin{figure}[!h]\begin{center}
\mbox{\epsfig{file=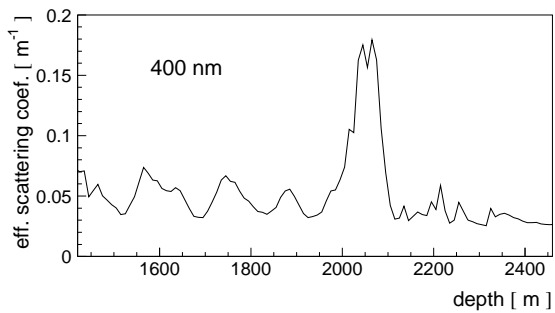,width=.55\textwidth}}
\parbox{.7\textwidth}{\caption[]{\label{sca}The effective scattering coefficientfor light at 400nm, the sensitivity maximum for IceCube DOMs, as a function of depth in the ice (from Ref.~\protect\cite{kurtice}).}}
\end{center}\end{figure}

The scattering and absorption values used are based on in situ measurements with AMANDA \cite{kurtice}.  This measurement was extrapolated to deeper ice using ice core 
data collected at Vostok station and Dome Fuji locations in Antarctica, 
scaled to the location of AMANDA using an age vs. 
depth relation \cite{agevsdepth}.  Figure \ref{sca} shows the resulting 
scattering coefficient as a function of depth. The scattering depth 
profile was further confirmed between 1300 and 2100 m
at much higher resolution with data collected by a dust measuring device (dust logger) used during the IceCube string deployment~\cite{dustlogger}.

For muon tracks, reconstructed parameters are the zenith angle, distance of the closest approach 
to the string, depth and time of closest approach to the string, 
and an estimate of the average photon density along the muon track.  The latter
is correlated with the average muon energy. 
Reconstruction of the azimuthal angle is not possible with a single string 
because of the cylindrical symmetry about the string.

\begin{figure}[!h]\begin{center}
\begin{tabular}{ccc}
\mbox{\epsfig{file=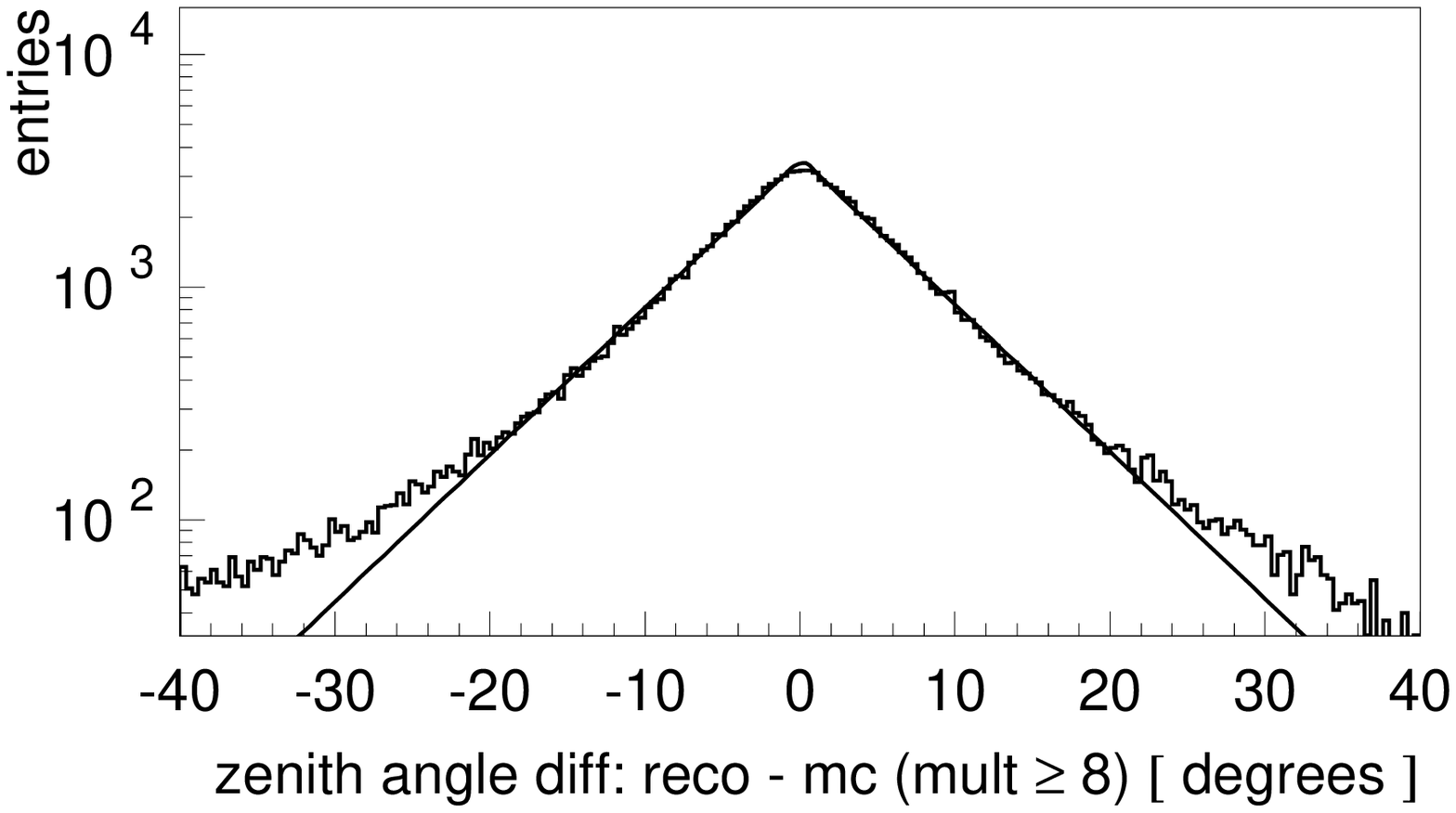,width=.44\textwidth}} & \ 
& \mbox{\epsfig{file=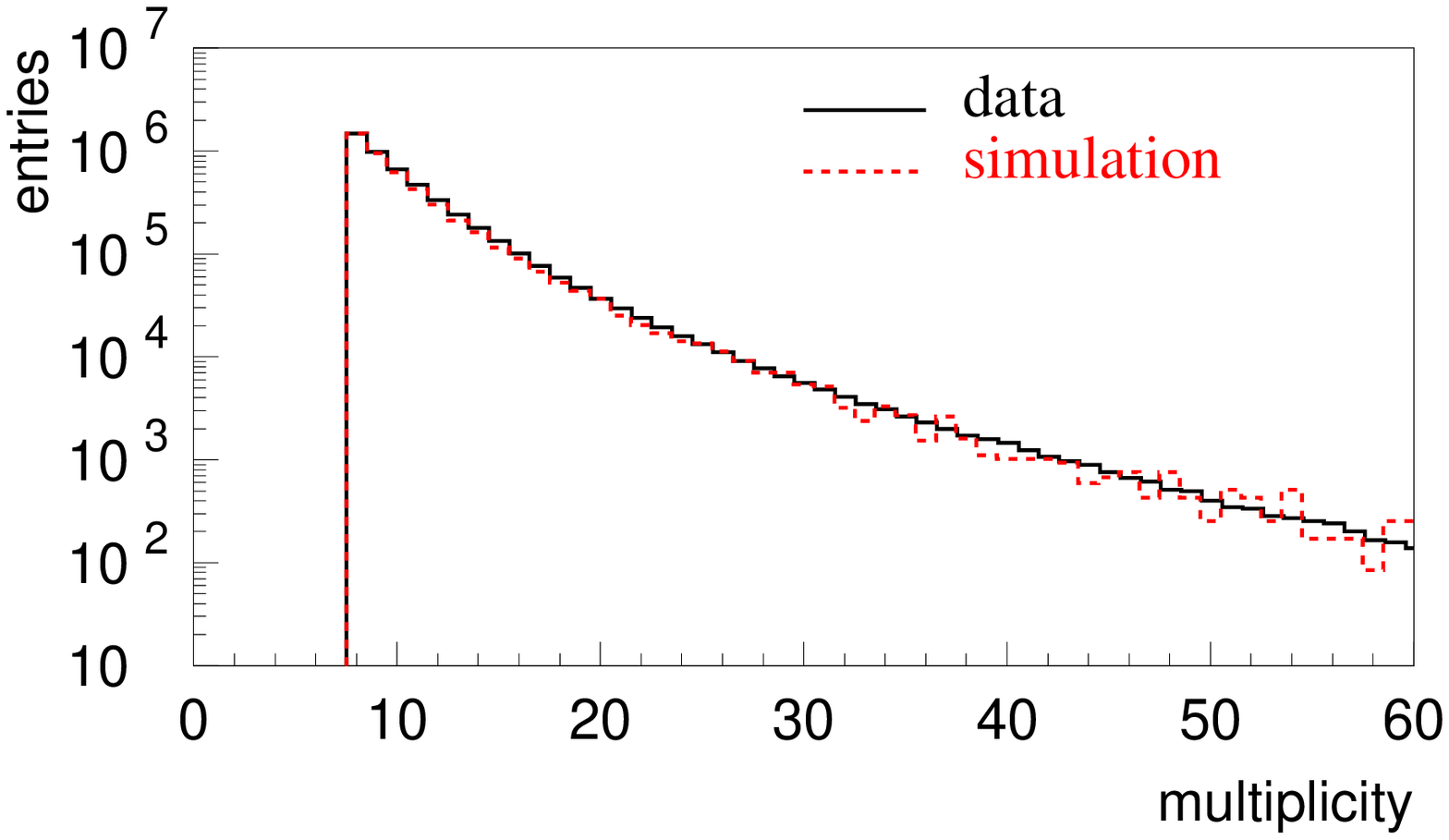,width=.46\textwidth}} \\
\parbox{.44\textwidth}{\caption[]{\label{a4} Zenith angle difference distribution of 
reconstructed and simulated tracks.}} & 
\ & \parbox{.46\textwidth}{\caption[]{\label{a5} Hit multiplicity distribution of 
muon events in data and simulation.}}
\end{tabular}
\end{center}\end{figure}

The track-fitting algorithm was tested on a simulated data sample of 
down-going muons and was found to reconstruct it rather well 
(given that only one string was used ~\cite{ICMuon}). 
Figure \ref{a4} compares the true zenith angle with the 
reconstructed zenith angle of simulated events.  The RMS resolution of the 
muon tracks with an event hit multiplicity of 8 or more is 9.7$^{\circ}$. 
The resolution improves rapidly as the multiplicity increases,
reaching ~1.5$^{\circ}$ for multiplicity of more than 30 hits.
This is similar to the one-string IceCube prototype analysis results \cite{DOM1}.
\begin{figure}[!h]\begin{center}
\mbox{\epsfig{file=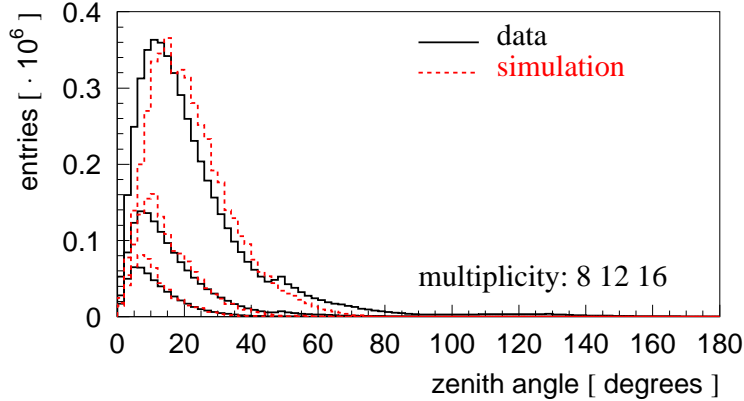,width=.70\textwidth}}
\parbox{.7\textwidth}{\caption[]{\label{a6}
Zenith angle distribution of reconstructed tracks in data and simulated muons.}}
\end{center}\end{figure}
%
\vspace{-0.1cm}

Figure \ref{a5} compares the hit multiplicity distribution for 8 hours of data and 
a similar amount of simulated data. 
The zenith angle distribution of the reconstructed 
tracks in data is compared to the simulated data in Fig. \ref{a6}. 
The simulated data 
used in Figs. \ref{a4}-\ref{a6} were produced with a standard AMANDA simulation,
which does not account for detailed differences in trigger logic, ice conditions 
and sensors of the deeper IceCube String-21.  As the IceCube simulation 
matures, the apparent discrepancy observed in Fig. \ref{a6} is expected to 
become smaller.
\subsection{Candidate Neutrino Search with a Single IceCube string}
About 99 \% of the muon data collected with the IceCube string
in 2005 was reconstructed and searched for 
upward-going muon candidates 
possibly induced by neutrinos.

A set of parameter cuts was applied to remove more uncertain, 
harder to reconstruct tracks, e.g., 
those that passed far from the string. 
These were satisfied by 70\% of reconstructed tracks,
corresponding to about 164 days of run time.

Additional cuts were developed to remove most of the remaining
misreconstructed tracks by rejecting events for which the 
most likely upgoing hypothesis was comparable to the most likely
downgoing hypothesis. 
\begin{figure}[!h]\begin{center}
\mbox{\epsfig{file=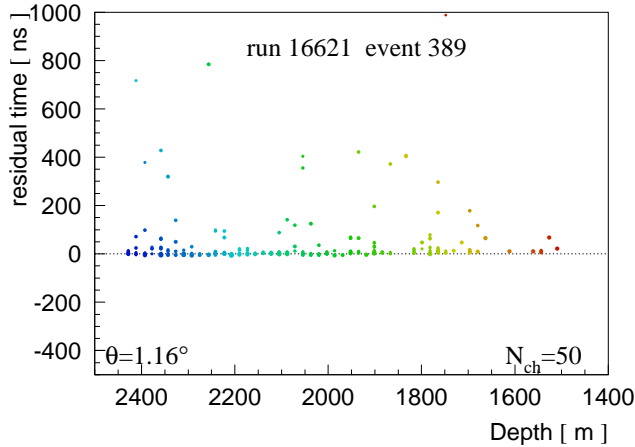,width=.60\textwidth}}
\parbox{.9\textwidth}{\caption[]{\label{nures}Photon arrival residual times for the 
event with multiplicity 50 reconstructed as upgoing muon track.}}
\end{center}\end{figure}

Two upgoing tracks were found with high hit multiplicity,
one with 35 and the other with 50 out of 60 DOMs hit.  These two events
are reconstructed as muons traveling up with 
nadir angles of 
1.16 $\pm$ 0.06$^{\circ}$ and 0.87 $\pm$ 0.05$^{\circ}$ degrees from the vertical, respectively. 
(The stated errors are statistical only; systematic errors
may be larger.)
Residual arrival times, i.e., time delays from reconstructed unscattered photon 
expectation, are plotted in Fig. \ref{nures} for one of these events. 
Almost every hit DOM received a direct (unscattered) hit, and many also recorded late 
(scattered) photons, more so in the dustier ice, and less in the clearer ice.
Observation of two nearly vertical upward muons with String-21 in
six months is consistent with expectations based on preliminary 
simulations \cite{IceCube}.

Several events with multiplicities ranging from 8 to 11 were reconstructed 
as traveling up at larger angles.  Confirmation of such events as upward-going
neutrino-induced muons would require more information, such as will be available from
adjacent strings in the future.

\subsection{Systematic timing offset study with downgoing muons}
\label{section:SystematicTimingOffset}
\begin{figure}[!h]\begin{center}
\begin{tabular}{ccc}
\mbox{\epsfig{file=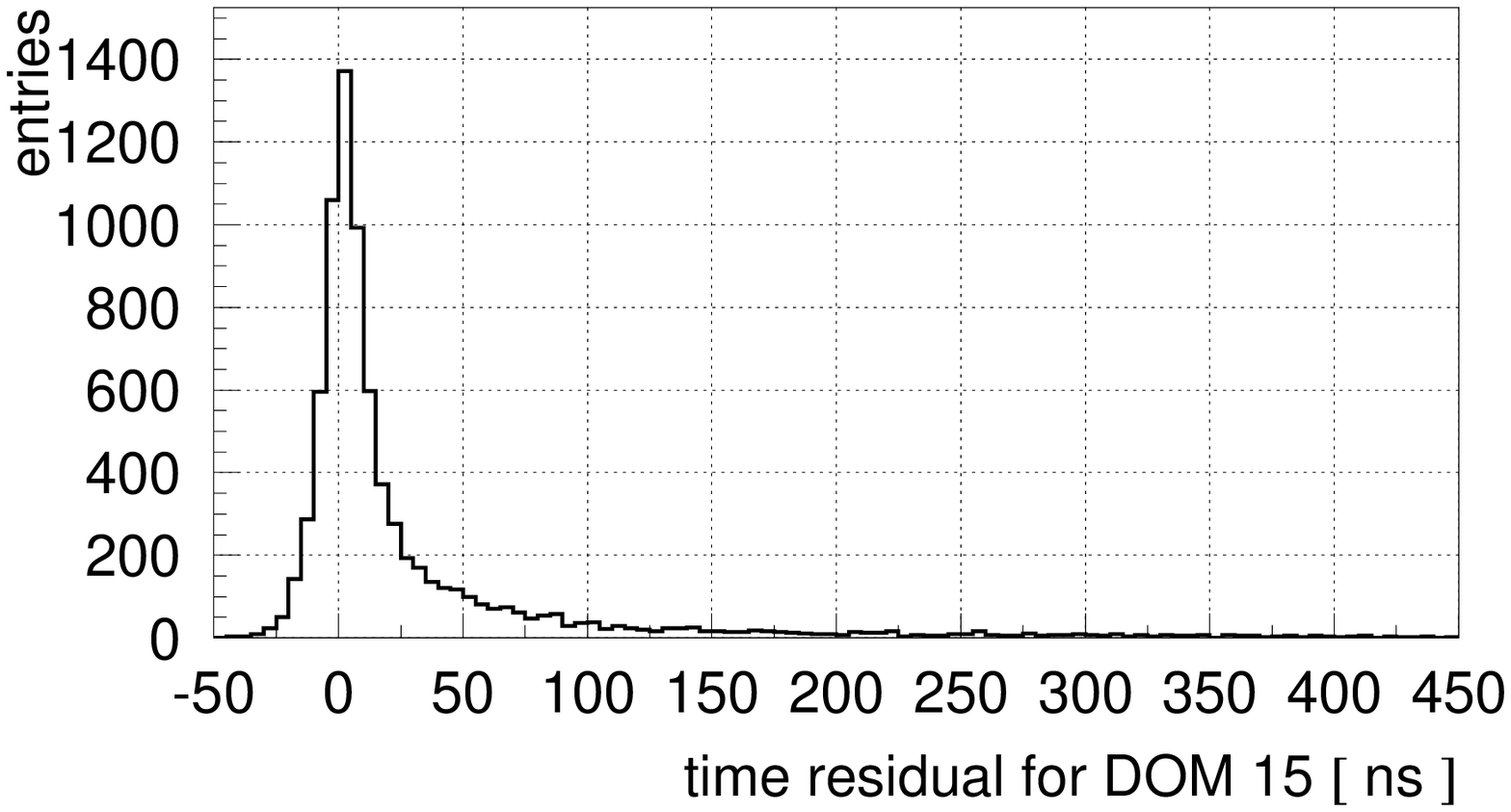,width=.45\textwidth}} & \ & \mbox{\epsfig{file=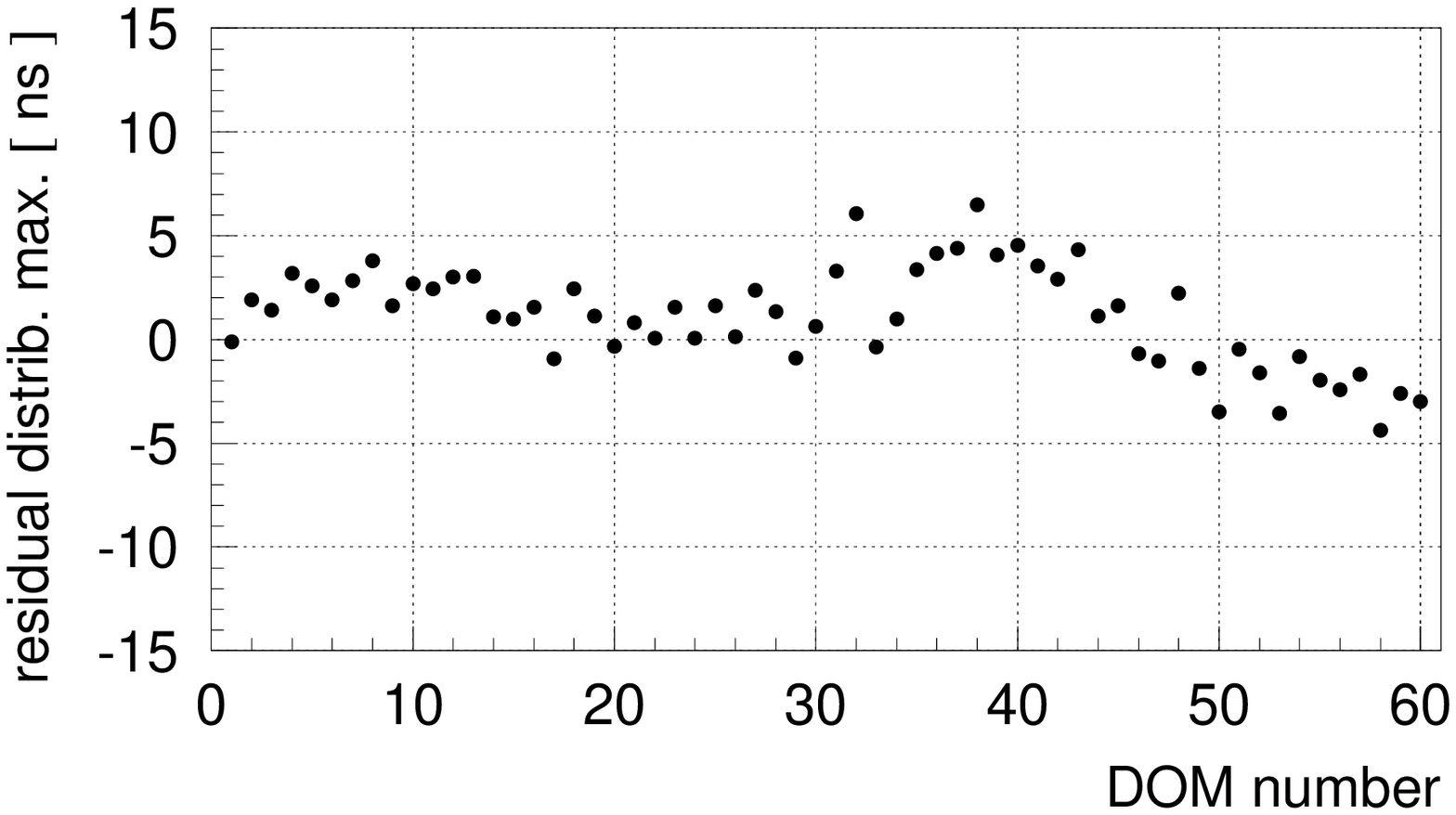,width=.45\textwidth}} \\
\parbox{.45\textwidth}{\caption[]{\label{a9} Distribution of time residuals of 
photons arriving at a DOM from nearby tracks reconstructed with the rest of the string. }} & \ & \parbox{.45\textwidth}{\caption[]{\label{aa} 
Plot of the peak of the distribution of time residuals for all DOMs on String-21.
(Figure \ref{a9} is the distribution for DOM 15.)}} \\
\end{tabular}
\end{center}\end{figure}

To measure systematic timing offsets in the IceCube string we repeated 60 times 
the single-string reconstruction  
for one day of data.
Each of the 60 DOMs was successively removed during the reconstruction,
and the time delays of the hits in each DOM with respect to the expected direct 
(unscattered) hit times from the reconstructed tracks (called time residuals) 
were evaluated. The residual time distributions are consistent with our expectation, 
with a narrow peak of nearly unscattered photons and a long tail of photons delayed 
by scattering, as illustrated for a particular DOM in Fig. \ref{a9}. 
The maxima of such distributions 
indicate the time residuals of the most probable hits. 

Systematic displacement of time residuals from zero could indicate errors in the 
time calibration procedure and possible variations in internal delays in different 
modules.  They could also be influenced by uncertainties in the DOM positions  
and by photon scattering in the ice.  
The peaks in residual distributions of most DOMs are within 3 ns of each other, 
i.e., the DOM clock times for the whole In-Ice array (currently 60 DOMs) are 
calibrated to within 3 ns of each other (with a mean of 1.1 ns and rms of 2.4 ns). 
DOMs 35-43 are located in dustier ice (the region between 2000 m and 2140 m 
in Fig. \ref{sca}) and have the highest residuals; DOMs 53-60 are located 
in the clearest ice (the region between 2300 m and 2430 m) and have the 
lowest residuals (Fig. \ref{aa}).  (Negative residuals are an artifact 
of
the fitting procedure.)
\section{Air-showers in IceCube}
\label{section:AirShowers}
As an air-shower detector, IceCube will have an unprecedented area for
detecting high-energy muons in the deep detector in coincidence with
the shower front at the surface.  Typical muons observed at
2 km depth have energies of 400 GeV or more at production
in the atmosphere.  The signal from the shower front at the surface
is primarily due to electrons and photons with energies below
1 GeV, with an additional contribution from muons in the GeV energy range.
The current four stations and one string are already
adequate to demonstrate the reconstruction of air-showers in IceCube.
\begin{figure}[htb]
 \includegraphics[width=7cm]{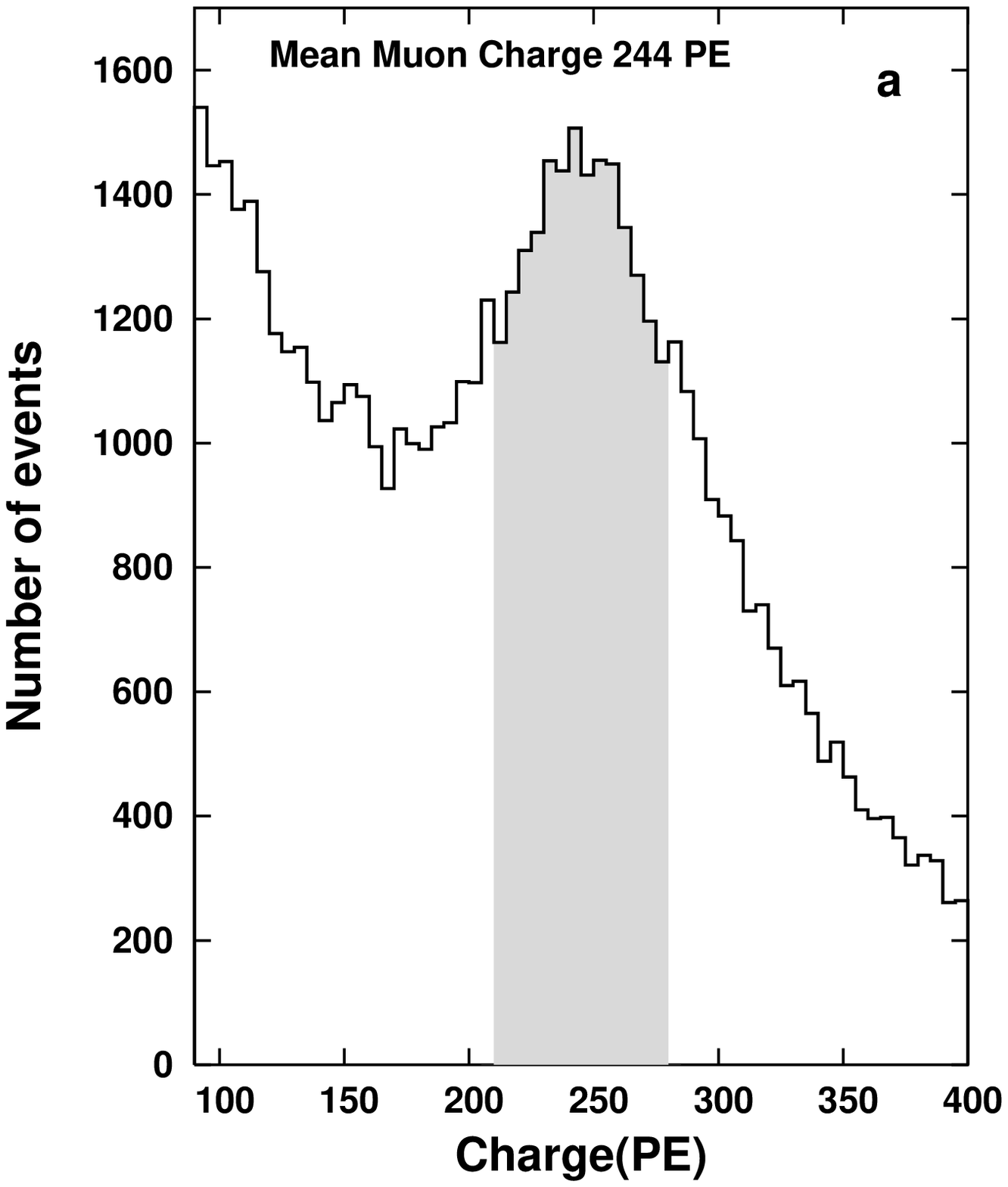}
\includegraphics[width=7cm]{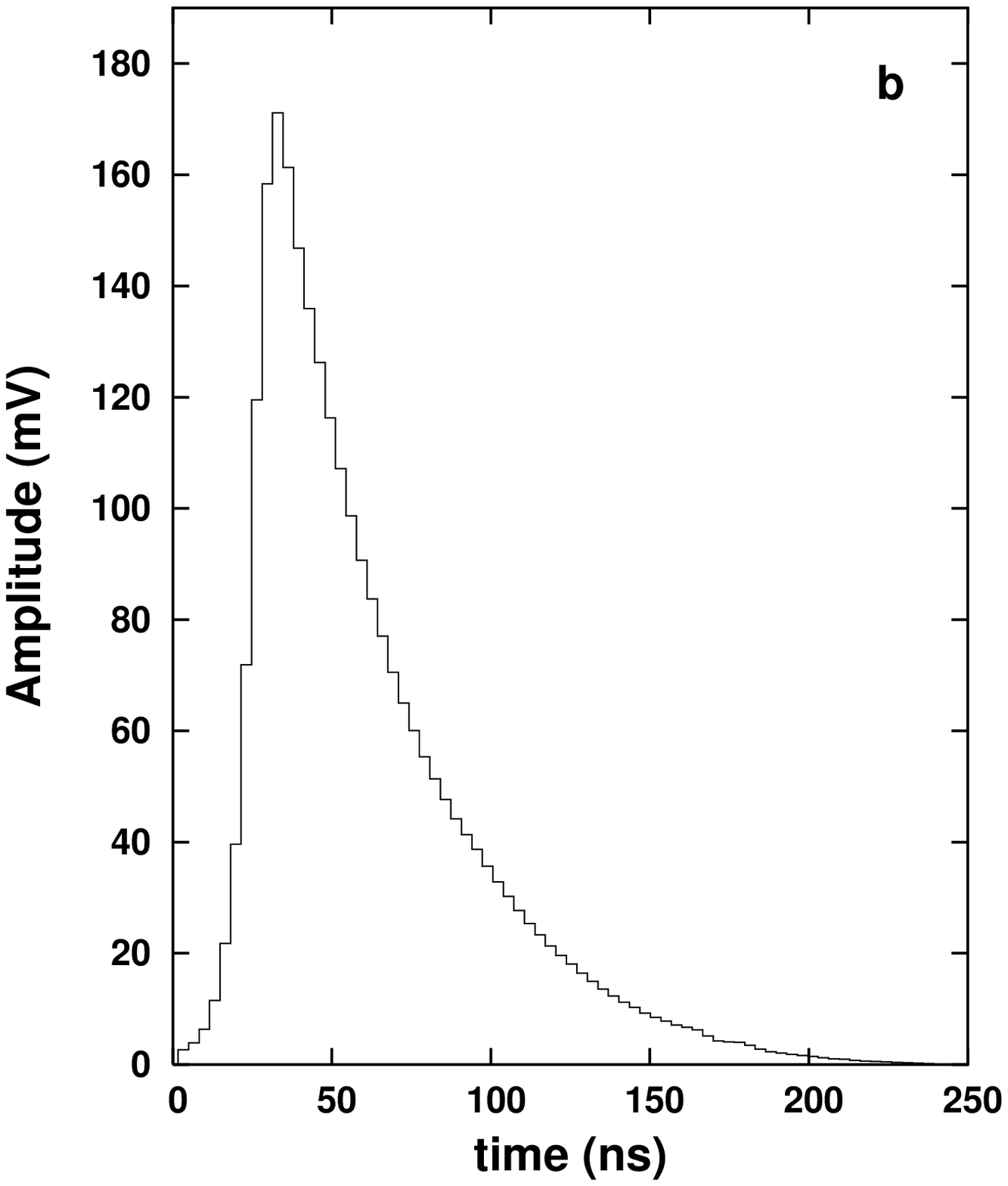}
\caption{Charge spectrum of signals in the high-gain DOM of tank A at station 39 (a) and
the corresponding average muon waveform (b).  These data were taken without any
coincidence requirement on the DOMs in order to obtain the spectrum of
muons, electrons and converting photons for calibration of each tank.
The shaded region indicates the muon peak.}
\label{mucalib}
\end{figure}
\subsection{Tank calibration and performance}
\label{section:TankCalibration}
Two types of data were taken with IceTop in 2005: calibration files and event
triggers.  For IceTop, the calibration files consist of tank events
with no coincidence requirement, as illustrated in Fig. ~\ref{mucalib}.
The figure shows the charge spectrum for the high-gain DOM
in one IceTop tank.  The spectrum for the low-gain DOM has
a similar shape.  The characteristic spectrum is a combination
of a steeply falling spectrum of electrons and $\gamma$-rays converting
in the tank with a peak due to muons traversing the tank.  There is also
a contribution from small air-showers, which becomes increasingly important
in the high-energy tail of the distribution.
A muon penetrating 90 cm of ice (the vertical height of the ice layer)
typically deposits approximately 190 MeV of energy in the detector
(more or less depending on its zenith angle).
The muon peak is broadened by the angular distribution of the muons,
by corner-clippers, and by fluctuations in energy deposition, as well
as by statistics of photon collection.  
The steady flux of
muons with its characteristic peak provides an ideal means of calibrating 
and monitoring a water
(or ice) Cherenkov detector~\cite{HP,Auger}.  

The single tank rate
corresponding to the threshold of 100 photo-electrons in
Fig. \ref{mucalib} is approximately 1.5 kHz.  Approximately 1 kHz of
this is from muons.  This rate is consistent with a previous measurement
using a muon telescope~\cite{Bai}.  
We use the peak of the inclusive muon distribution at approximately
240 PE for calibration.  Each PE corresponds to a nominal energy deposition
of $$ {190 {\rm MeV} \over 240 \rm PE}\;\approx\;0.8 {\rm\; MeV/PE}.$$
\begin{figure}[htb]
 \includegraphics[width=14cm]{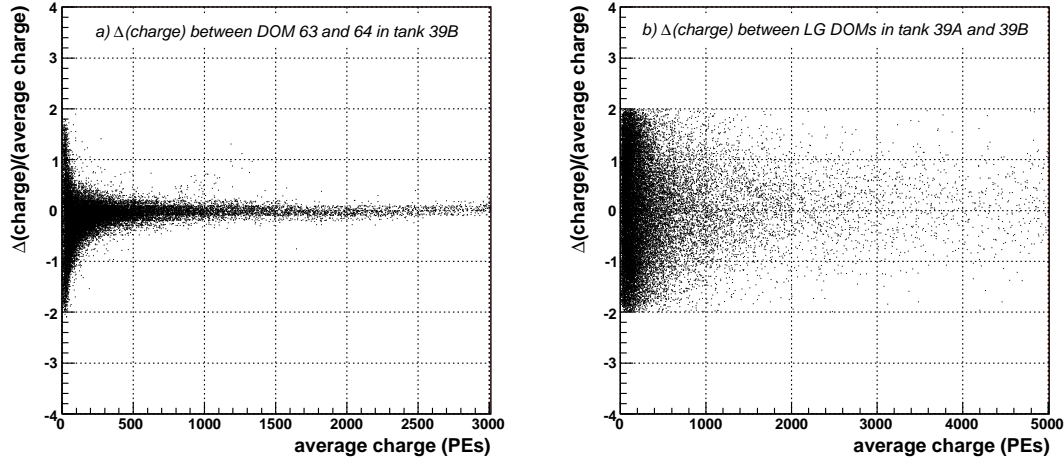}
\caption{(a)Fluctuations in tank response to signals
measured by comparing the response of two DOMs
in the same tank to air-showers. (b) Shower-front
fluctuations measured by comparing response of two tanks at the
same station to air-showers.
}
\label{tanksAA}
\end{figure}

We can assess the uniformity of response of a tank to a given
deposition of energy by comparing the response of the two DOMs
in the tank to the same events.
Differences in
response can occur for several reasons, including small deposits of
energy farther
from one DOM than the other, direct hits on the photocathode, etc.
The in-tank fluctuations in response are measured using one high-gain
and one low-gain DOM, so this study is limited to the range of common
linear response.  The result is shown in Fig. \ref{tanksAA}a for a sample of 
air-showers.  (Both panels show difference in charge divided by average charge,
so the results are confined by definition to lie between $-2$ and $+2$.)

To decide whether the tank response is sufficiently uniform,
we compare the response of two tanks
at the same station to the same set of air-shower events in Fig.~\ref{tanksAA}b
using the low-gain DOM in each tank.
To the extent that fluctuations in single tank response are small
compared to differences between tanks, this is a measure of
fluctuations in the air-shower front.  This result shows that
uniformity of tank response is satisfactory because fluctuations in
response of a tank to a particular deposition of energy are small
compared to intrinsic fluctuations in the shower front.
\subsection{Shower reconstruction}
\label{section:ShowerReconstruction}
Reconstruction of the direction and size of an air-shower proceeds
from the arrival times of the shower front at the detectors on the
ground together with a measure of the number of particles or amount of
energy deposited in each detector.  
As evident from Fig. \ref{tanksAA},
typical air-shower signals produce hundreds or thousands of
photoelectrons in IceTop tanks.  Some examples of the resulting
waveforms are shown in Fig. \ref{waveform}.  
The waveforms occasionally show structure that
may reflect contributions of individual particles or groups of
particles.

Several algorithms were developed to estimate the arrival time of the
first particle in the shower front at an IceTop tank (leading edge).
The most robust algorithm looks for the first pair of bins in the
waveform with values above a fixed threshold between which the
increment in voltage is locally at maximum (i.e. the steepest rise point).  
The intersection of the
tangential line going through these points with the baseline is taken
as an estimate of the leading edge. The sum of charges in all bins of
the waveform above the threshold is taken as an estimate of total
charge in the waveform.  
The time of the leading edge of the waveform
can be determined to significantly better accuracy than the 3.3 nsec sampling rate of the ATWD.  
\begin{figure}[htb]
 \includegraphics[width=7.cm]{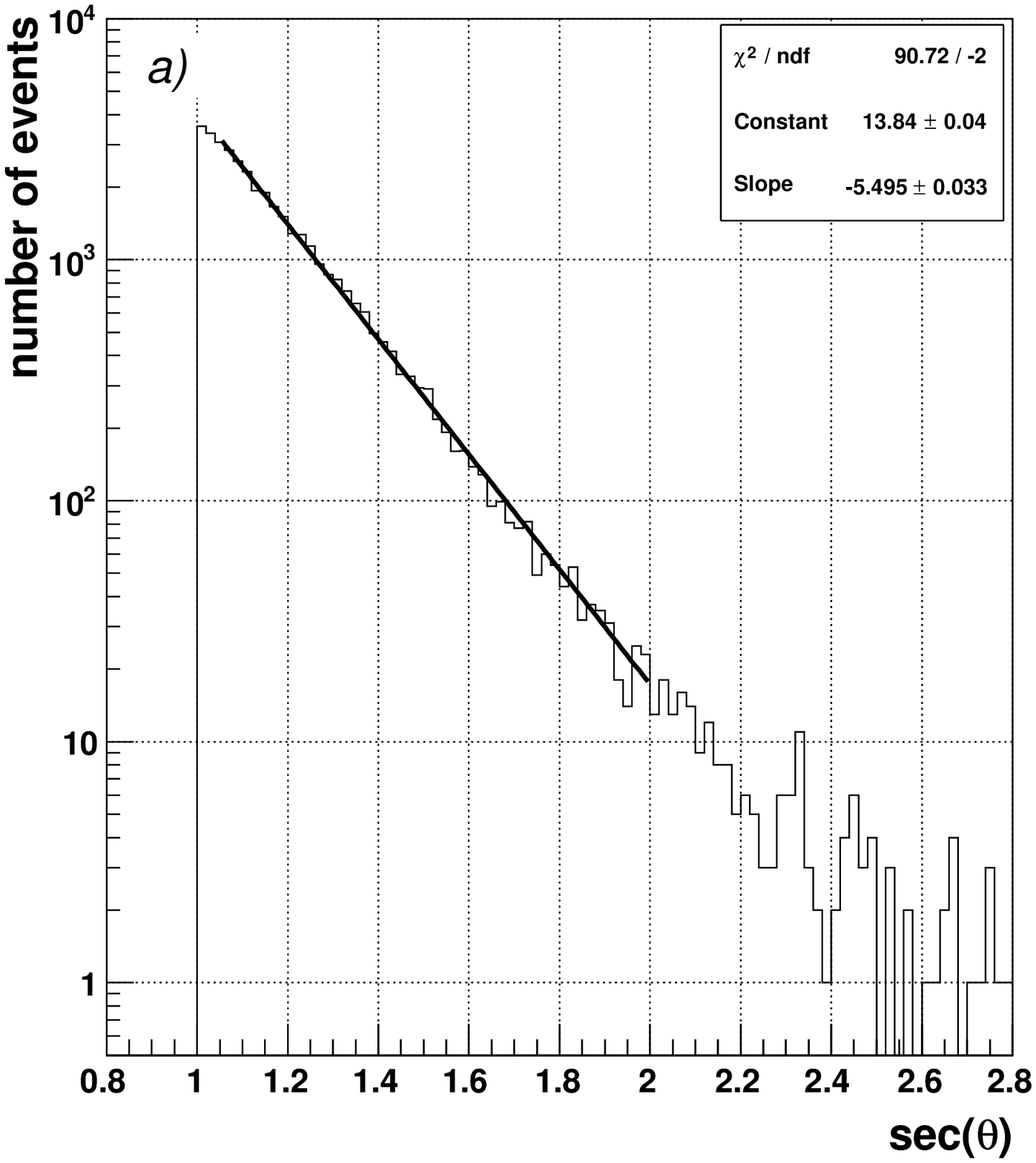}
 \includegraphics[width=7.cm]{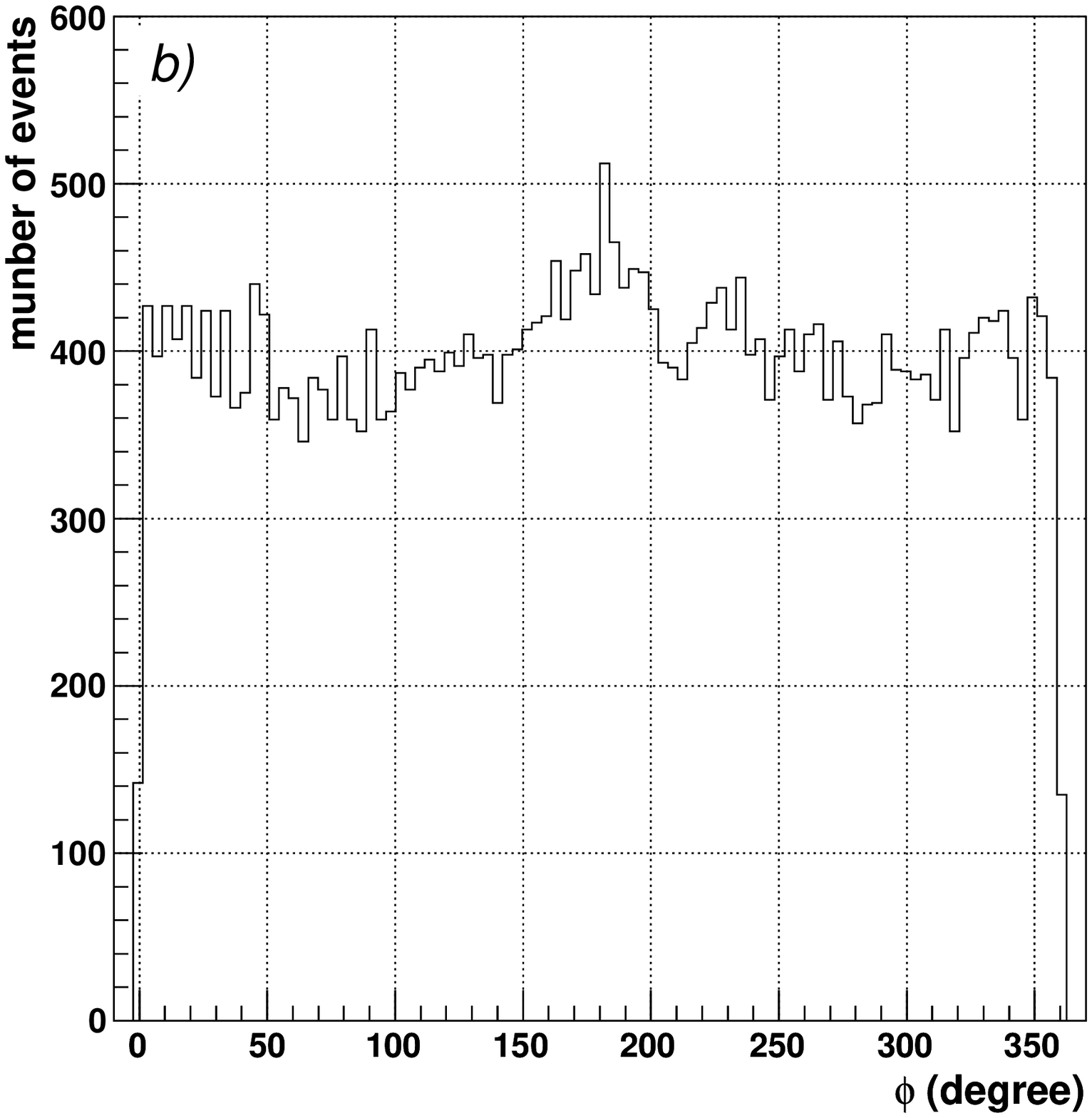}
\caption{Reconstructed directions of showers measured
with the surface array: (a) zenith angle. (A fit to the sec$\theta$ 
law typical of atmospheric showers is shown.) (b) azimuth. 
}
\label{angles}
\end{figure}

From February to July 2005 the IceTop trigger required 10 DOMs within 
2 $\mu$sec with
signals above a voltage threshold equal to ten times the peak voltage
of a single photoelectron.  The trigger rate with this setting is approximately
0.7 Hz, about 40\% of which involve all four stations, while
the remainder have hits in only three stations.  With only four
stations, many of the triggers are from showers with cores outside the
perimeter of the array, where the core location accuracy is poor.  
For the initial analysis we fit a plane wave to the shower front.
By selecting showers with apparent
cores (as determined by weighting the tank locations by the observed
signals) within 45 m of the geometric center of the 4-station array,
we obtain a subset enriched in ``contained'' events.  
Distributions of
zenith and azimuth for this subset are shown in Fig. \ref{angles}.

To obtain the most accurate possible
determination of shower direction requires accounting for the delay of
the leading edge behind a plane perpendicular to the trajectory of the
cascade, which increases with core distance.  Reconstruction accuracy
is also limited by distribution of arrival times of the first particle
in the shower front, which depends on shower size and distance from
the shower core, and on accuracy of location of the core.  More refined 
fits to shower direction and core location will become appropriate in
future seasons when the array is larger.
\begin{figure}[htb]
\includegraphics[width=7cm]{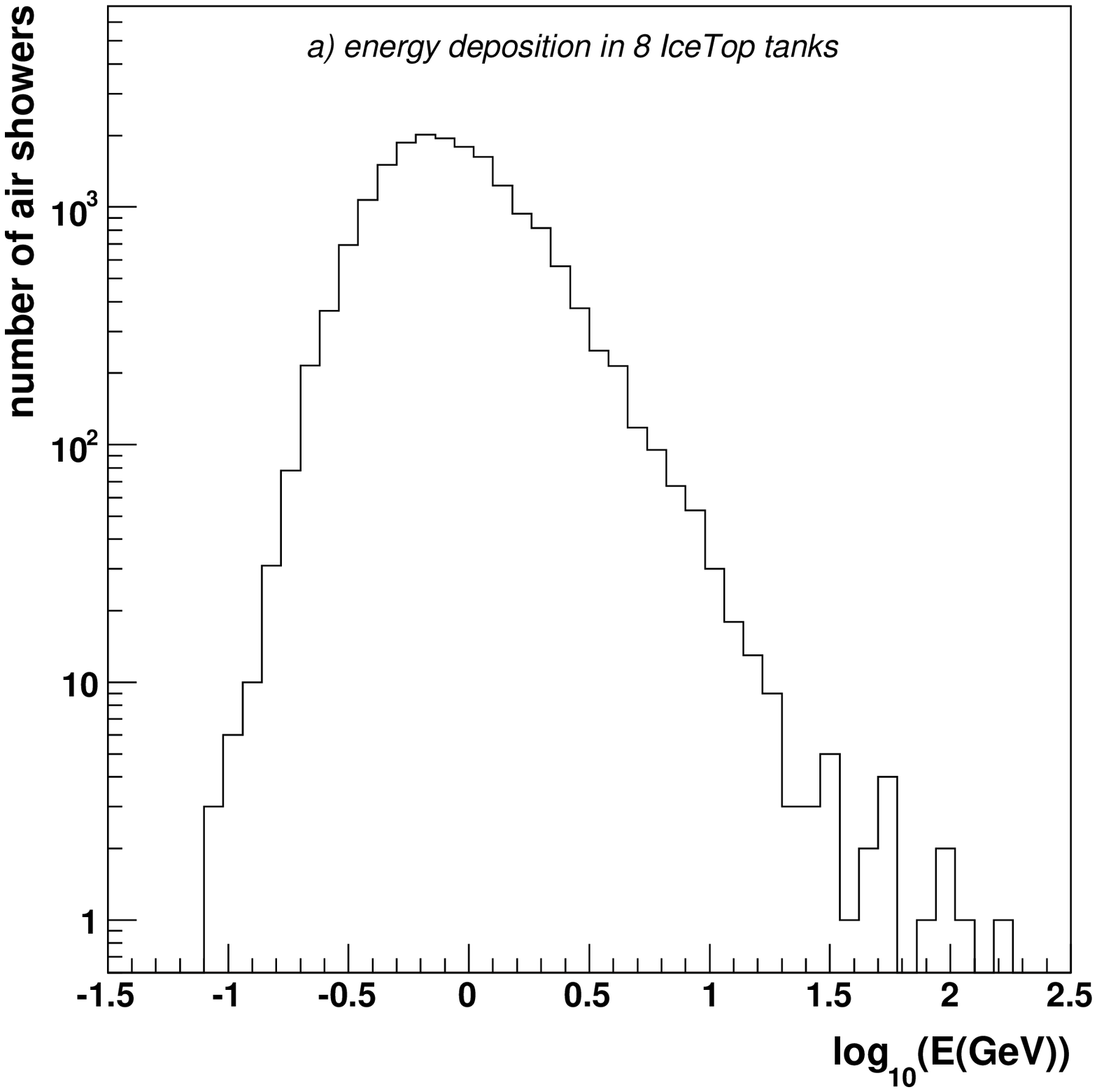}
\includegraphics[width=7cm]{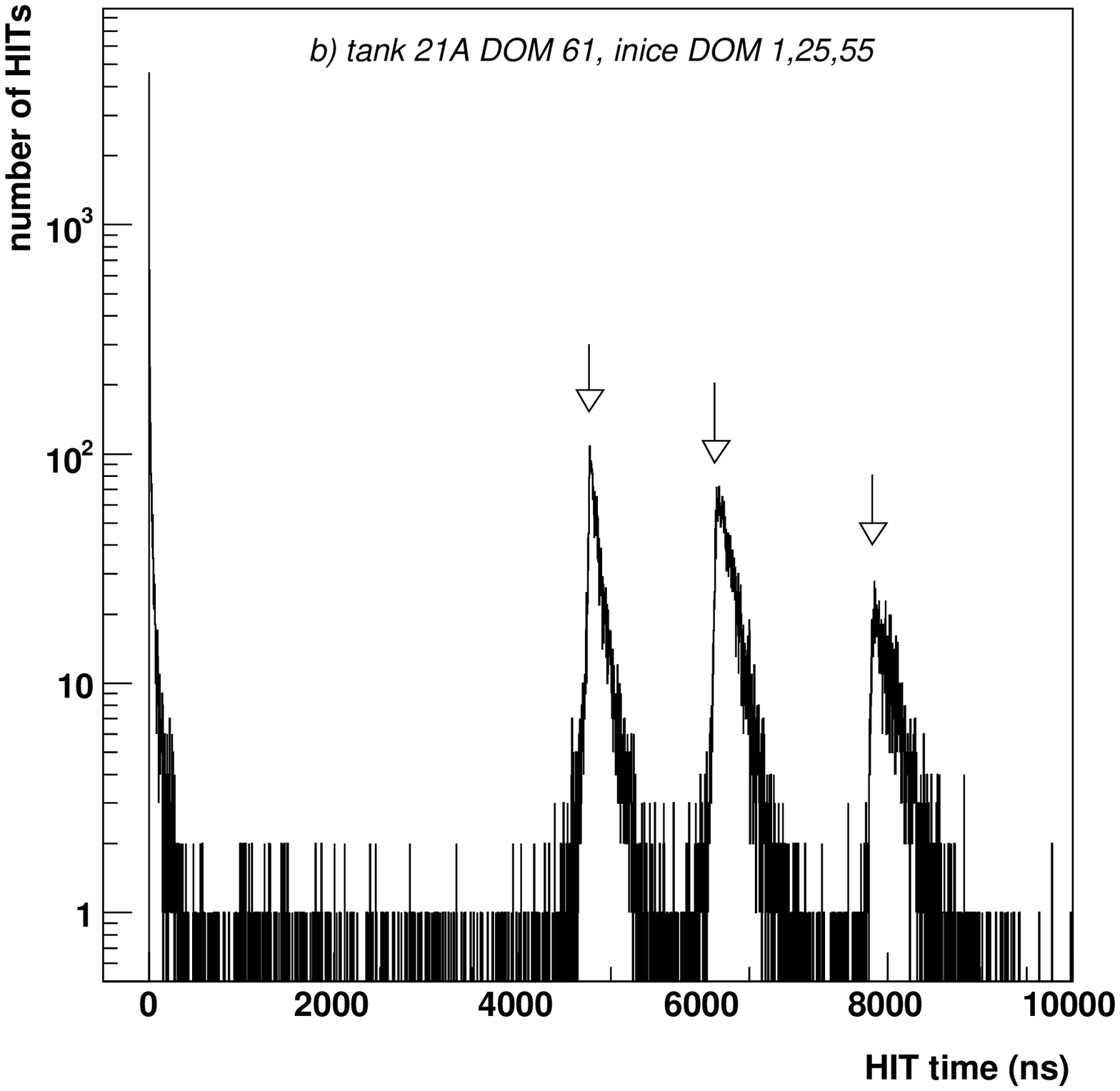}
\caption{(a) Shower-size spectrum in units equivalent
to total visible energy deposited in the 8 tanks; (b)
Times of 3 specific DOMs on String-21 in coincident events
relative to the first IceTop DOM in the event.
The peak near 0 is the time a specific IceTop DOM is hit.}
\label{spectrum0}
\end{figure}

Figure \ref{spectrum0}a shows the size-spectrum of showers measured by
the IceTop array for a sample of events with apparent core location
within 45 m of the center of the array.  Shower size at the ground is
measured here in terms of total visible energy summed over all tanks,
based on the calibration with muons discussed above.

Studies of the cosmic-ray energy spectrum and composition require
detailed simulations which are not yet available.
A crude estimate using the ratio of the total area of the array to the
total area of the 8 tanks leads to an estimate of total $N_\mu\,\sim
4\times 10^4$ at 10 GeV visible energy, which corresponds to showers
with primary energy of 1 PeV.  The visible energy spectrum of
Fig. \ref{spectrum0}a has a power low behavior at high energy and a
characteristic threshold shape at low energy.  Given the spacing of
the stations, IceTop with a 4-station threshold is expected to be
50\% efficient at an energy of $300$~TeV.  This is consistent with the
measurement shown in Fig. \ref{spectrum0}a, which includes 3-station
hits as well as 4-station hits.
\subsection{Single station coincidences with String-21}
\label{section:SingleStationCoincidence}
Air-showers containing one or more muons with energy sufficient to
reach the deep detector can be detected in
coincidence by IceTop and String-21.  Because of the steep cosmic-ray
spectrum, most of these coincident events will produce signals at only
one IceTop station.  In a small fraction three or more IceTop stations
will be hit, satisfying the IceTop trigger described above.  Events in
which both tanks at one station are hit with no hits in any other
stations are highly correlated with single muons at string depth and
therefore of special interest for calibration and background tagging. 
The rate of single-station coincidences with String-21 triggers is
0.05 Hz.  By comparison with a preliminary simulation, this rate is
consistent with expectation if String-21 sees vertical muons out to 30
m with its present 8-fold coincidence requirement; this 30 m radius is
consistent with the measured In-Ice muon rate.  As the array grows, the rate
of such tagged single muons will increase quadratically, allowing a large
sample of tagged background events to be accumulated.  It will be possible
to tag about 5\% of the In-Ice muons, including some accidental coincidences
within a 2 $\mu$sec trigger window. 

Because of the large spacing between the DOMs in IceTop and those in
String-21, the coincident events allow us to verify the absolute
timing of IceCube on the kilometer scale, using a selection of tagged,
nearly vertical muons.  Fig. \ref{spectrum0}b is a distribution of
times for events in which at least one of 3 specific DOMs on String-21
(1, 25 and 55) is hit in coincidence with a specific high gain DOM in
a surface tank.  (DOM 1 is at the top of String-21 and DOM 60 at the
bottom.)  Times on String-21 are measured relative to the time of the
first signal in IceTop (which may occasionally be earlier than the
time of the specific IceTop DOM plotted).  By definition, there are at
most 4 entries for each event.  The systematic decrease in population
of the deeper DOMs on String-21 is an expected consequence of muons
ranging out between the top and bottom of the string.
\subsection{Air-showers in coincidence with String-21}
\label{section:AirShowersCoincidence}
Showers with 3 or 4 stations hit contain sufficient information so
their directions can be reconstructed independently of String-21.
The rate of such events is $0.002$~Hz, approximately $0.3$\% of the
total rate of air-showers in IceTop.  Given the angular distribution
shown in Fig. \ref{angles}a, this fraction corresponds to detection of
all trajectories that pass within approximately 60 meters of String-21, 
which is comparable to the size of the muon bundles at a depth of $2000$~m.
Comparison between the direction assigned to the same events
independently by IceTop and by String-21 (zenith angle only) can 
be used to calibrate the pointing and angular resolution of the deep detector.

\begin{figure}[!h]\begin{center}
\begin{tabular}{ccc}
\mbox{\epsfig{file=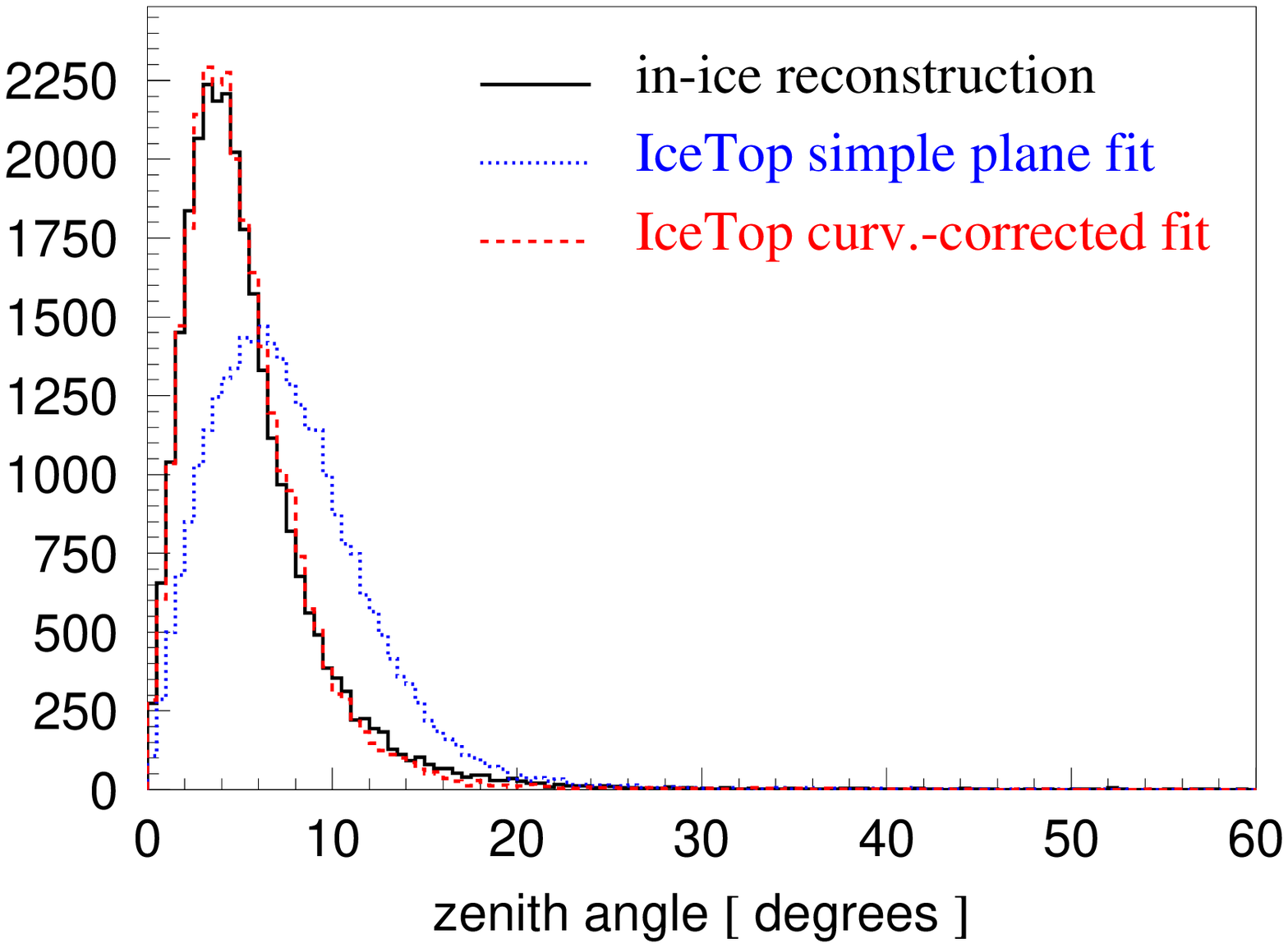,width=.45\textwidth}} & \ & \mbox{\epsfig{file=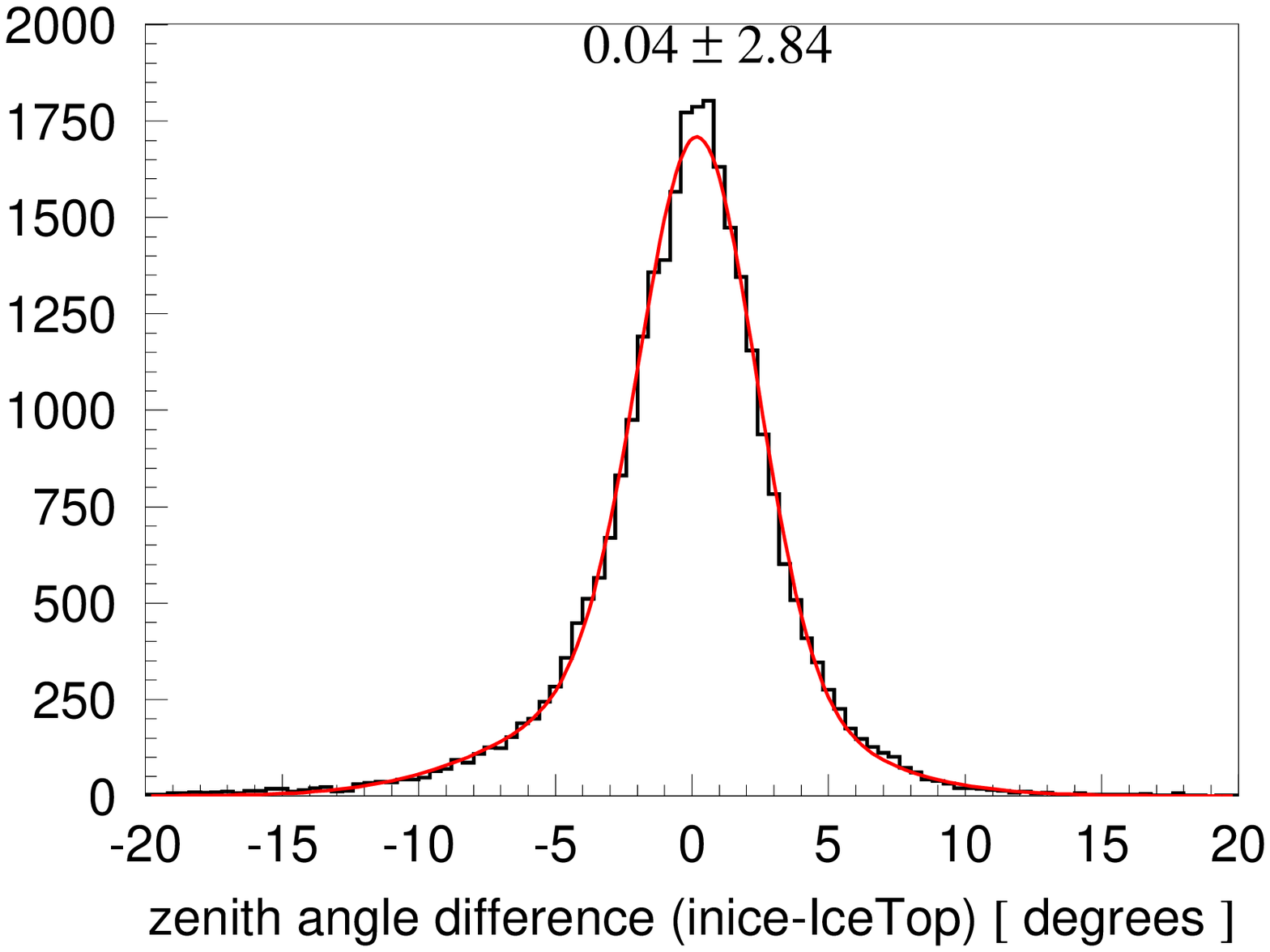,width=.45\textwidth}} \\
\parbox{.45\textwidth}{\caption[]{\label{a7} Zenith angle distribution of string-reconstructed tracks (solid) and IceTop-reconstructed coincident showers (dotted for simple plane $\chi^2$ fit and dashed for curvature-corrected likelihood reconstruction). }} & \ & \parbox{.45\textwidth}{\caption[]{\label{a8} Zenith angle difference distribution between string-reconstructed tracks and IceTop-reconstructed coincident showers. }} \\
\end{tabular}
\end{center}\end{figure}

To explore this approach, we selected coincident In-Ice and IceTop
events with a combined hit multiplicity of at least 8 In-Ice +10 IceTop hits
collected during the year 2005.  Selected showers were
reconstructed with both the IceTop shower reconstruction and the
one-string muon track reconstruction discussed in section \ref{section:MuonReconstruction}.  The resulting
zenith angle distributions are compared in Fig. \ref{a7}.  Directions
obtained with the string reconstruction are systematically closer to
the vertical than those obtained with a simple plane fit algorithm to the IceTop showers. Using the shower curvature parameterization of \cite{hinton}, obtained for SPASE (located nearby at the same atmospheric depth and having similar dimensions to the 4-station IceTop array) brings string and IceTop reconstruction results in agreement as shown in Fig. \ref{a7} and Fig. \ref{a8}.
\section{Coincidences of String-21 with AMANDA and SPASE}
\label{section:CoincidenceStrin21AmandaSPASE}
\begin{figure}[htb]
\begin{center}
\includegraphics[width=12cm]{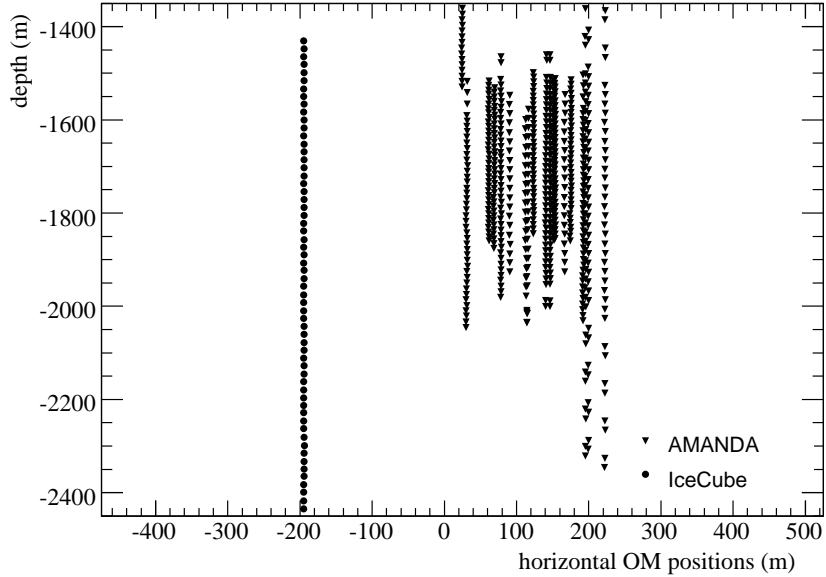}
\end{center}
\caption{Geometry of AMANDA and String-21. }
\label{geom}
\end{figure}

Coincidences with two existing nearby detectors allow further exploration
of the response of the IceCube string.  The center of AMANDA
is 328 m from String-21, and most AMANDA modules are between
1500 and 2000 m deep, as shown in Fig. \ref{geom}.  The center
of SPASE is 248 m from the top of String-21, so trajectories
through SPASE and String-21 have zenith angles between $5^\circ$
and $10^\circ$.
  
When these data were taken, neither AMANDA nor SPASE was directly 
connected to IceCube. As a consequence, coincidences with these detectors were found using the event times from their independent GPS clocks. In the January 2006, the SPASE, AMANDA, and IceCube triggers were connected to permit coincident triggering, and a system for synchronizing the reference times of the three detectors was installed. The detectors will be integrated further in coming years.
\subsection{String-21 and AMANDA}
\label{section:String21Amanda}
\begin{figure}[htb]
\includegraphics[width=7cm]{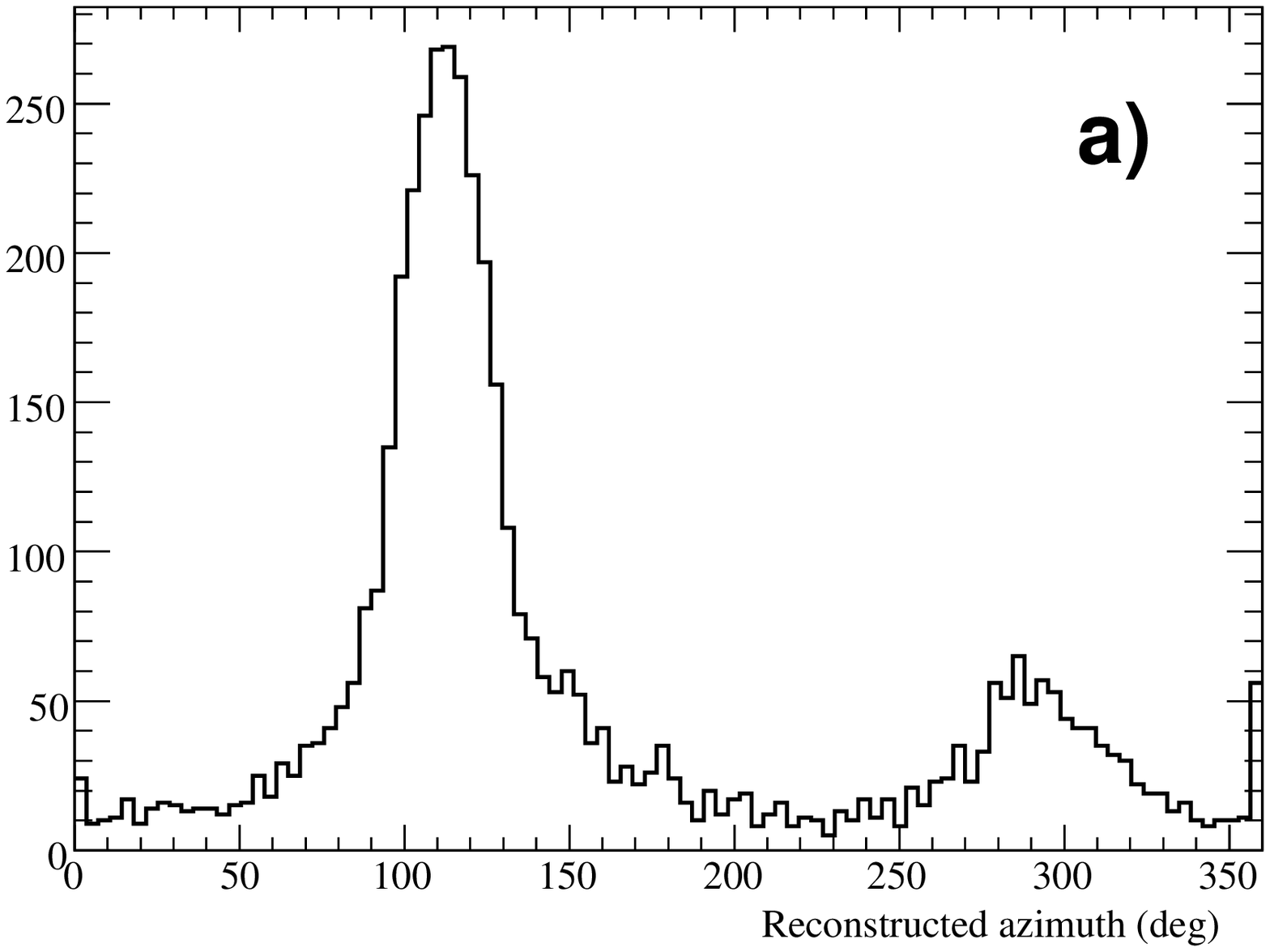}
\includegraphics[width=7cm]{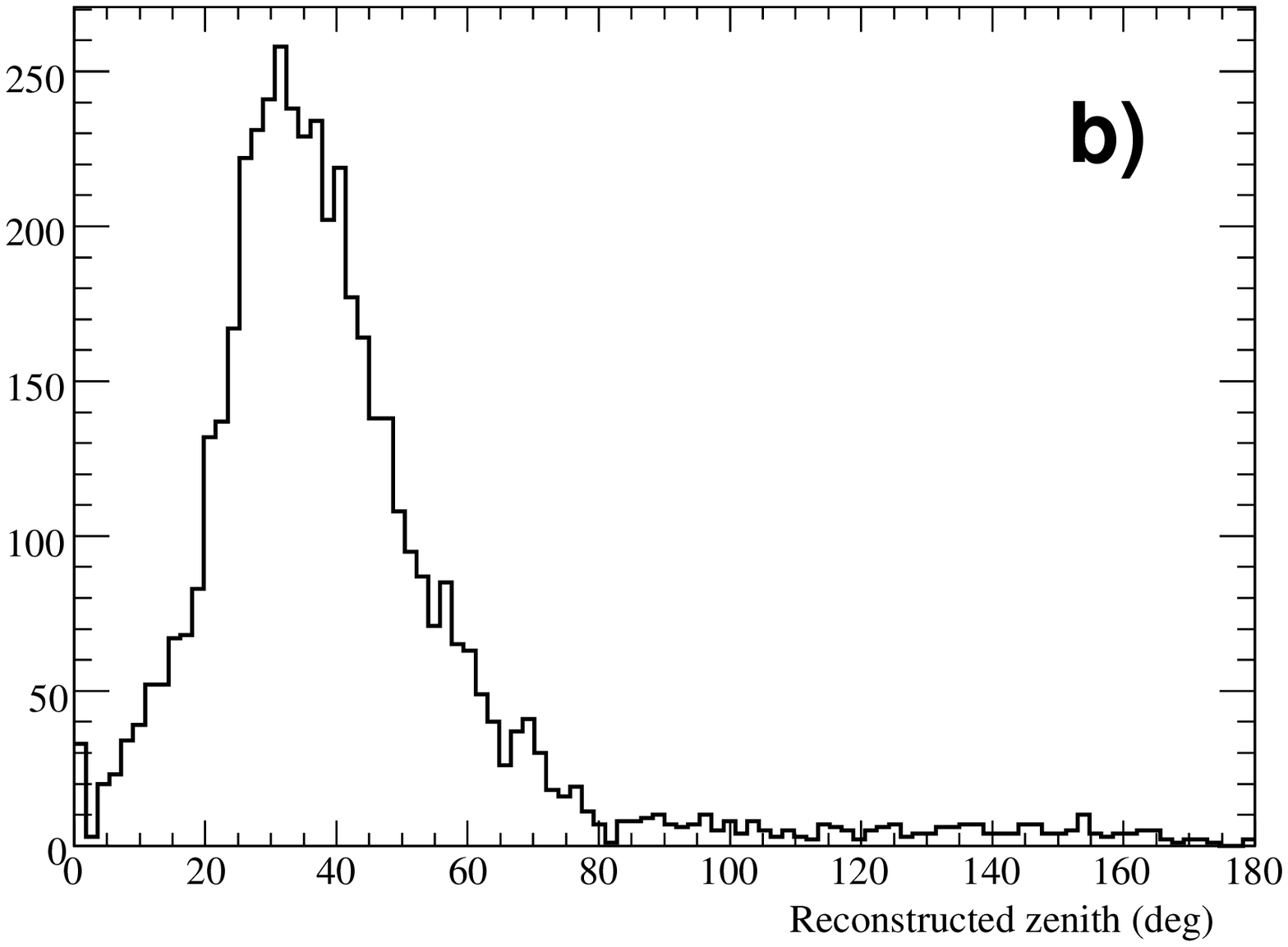}
\caption{Angular distribution from AMANDA for String-21
coincident events: Azimuth(a) and Zenith(b).}
\label{A-21-angles}
\end{figure}
\begin{figure}[htb]
\begin{center}
\includegraphics[width=12cm]{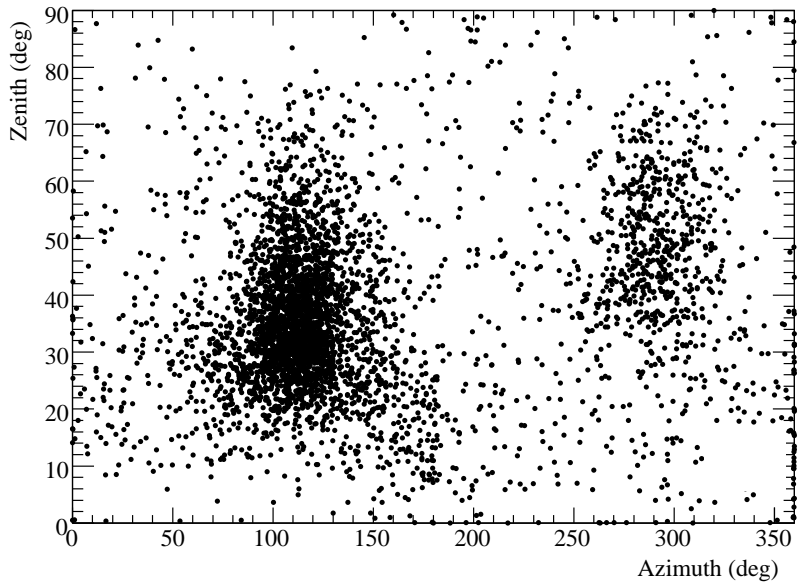}
\end{center}
\caption{ Scatter plot of directions of AMANDA-String21 coincidences. }
\label{scatter-amanda}
\end{figure}

Figure~\ref{A-21-angles} shows the zenith and azimuthal distributions
for AMANDA coincidences with String-21.  There are two peaks in the
azimuthal distribution separated by 180$^\circ$ depending on which
detector the muon hits first.  The relative population of the two
peaks is determined by the convolution of the geometry of the two
detectors with the steeply peaked zenith angle distribution of
atmospheric muons.  IceCube is deeper than AMANDA, so most muons go through AMANDA first.  The corresponding shift in the
zenith angle distribution is apparent in the scatter plot, which
is shown in Fig. \ref{scatter-amanda}.

\subsection{String-21 and SPASE}
\label{section:String21SPASE}
\vspace{-.5cm}
\begin{figure}[htb]
 \includegraphics[width=7cm]{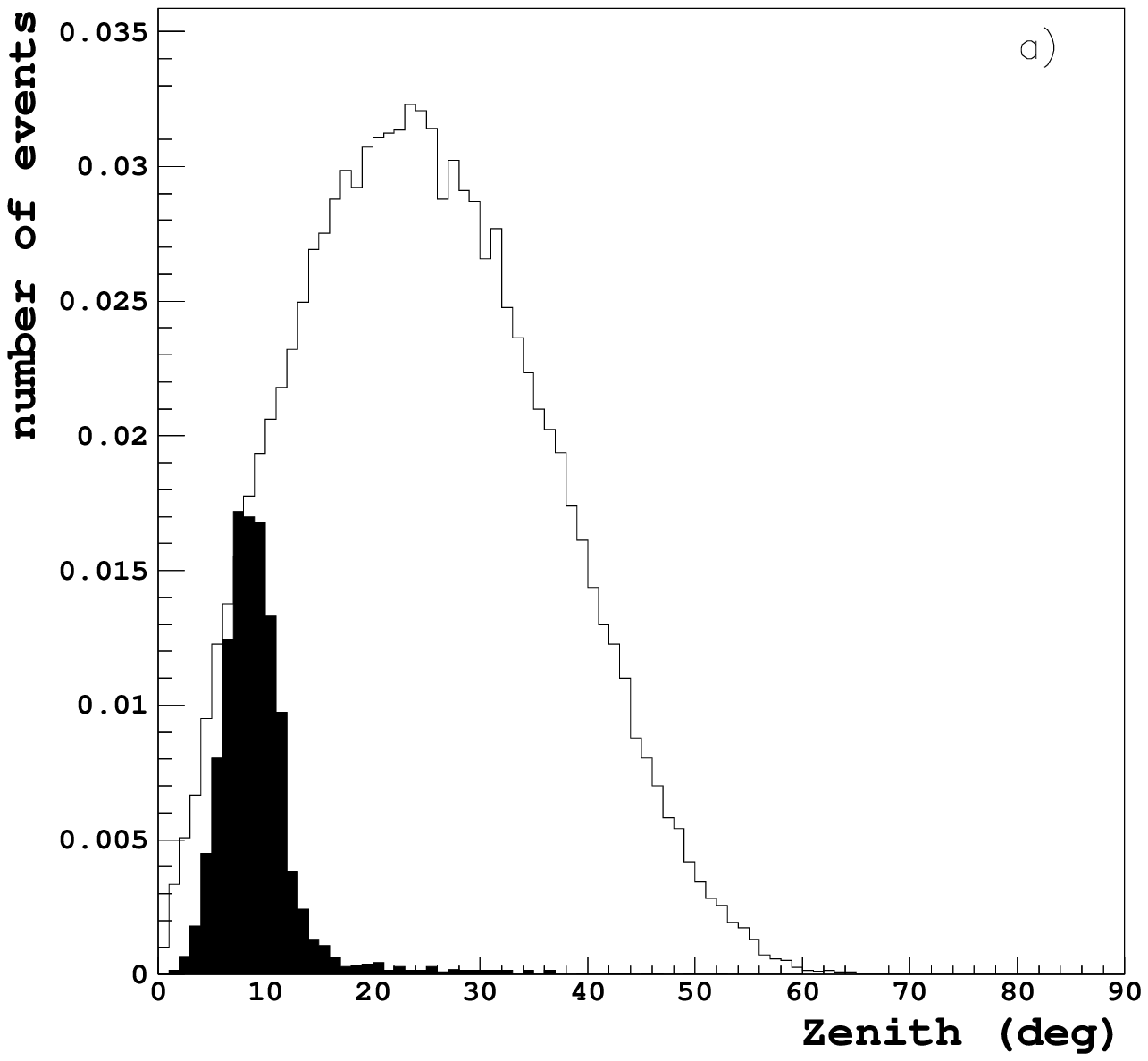}\hfill
 \includegraphics[width=7cm]{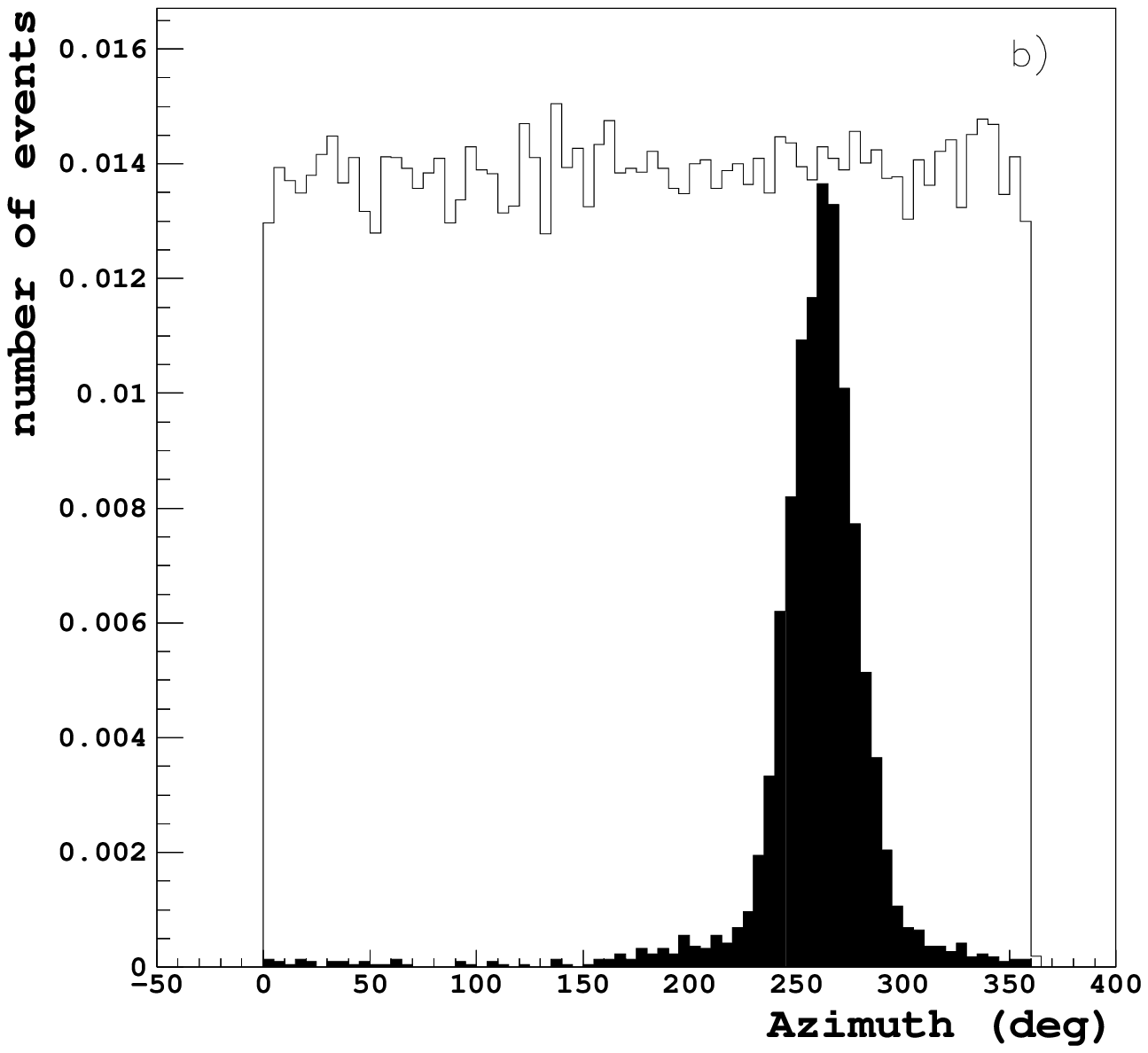}
\caption{
SPASE and String-21: Distribution of zenith (a) and
azimuth (b) of showers reconstructed by SPASE.  The black histograms
show events with a muon trigger in String-21, while the unshaded
histograms show all showers reconstructed by SPASE.  Azimuth
is defined here in the SPASE coordinate system. 
}
\label{spst21}
\end{figure}
The SPASE scintillator array has a detector spacing of 30 meters and
hence a threshold for air-showers about an order of magnitude lower
than IceTop.  It can therefore extend the IceCube
primary composition analysis to lower energy.  The SPASE data
acquisition system was upgraded at the end of 2002, and analysis of
coincidences with AMANDA-II is currently underway to extend
the investigation carried out with AMANDA-B10 before 2002~\cite{SPAM2}.  
Fig. \ref{spst21} shows the showers that
are seen by both SPASE and String-21.
\section{Summary and Plan for Completion}
\label{section:Summary}
We have deployed and operated one string and eight surface detectors
of the IceCube Neutrino Observatory in 2005.  The separation between
the digital optical modules on the surface and those near the bottom of the
string in the deep ice is 2.5 km, comparable to the scale
projected for the finished IceCube.  Successful reconstruction of 
coincident cosmic-ray events with few nanosecond accuracy therefore
demonstrates that the full detector will work as planned.

All 76 PMTs in the DOMs are working.  Remote
calibration of the timing and amplitude of the digital optical modules
is carried out successfully as part of the normal operation
of the detector.  Tests with downward cosmic-ray muons and with
artificial flashers demonstrate that timing accuracy at the level 
of a few nanoseconds is maintained over all the DOMs.
These accurate calibrations have allowed the reconstruction of the first upward-going
candidate neutrino induced events with unambiguous time patterns on more than 30 DOMs. The zenith angle 
distribution of atmospheric muons has been reconstructed with String-21 
and the time residuals for each DOM have been measured using muons.
The ability to reconstruct waveforms with few nanosecond resolution
will be a powerful tool for analysis of complex events.  

The presence of a surface array above a neutrino telescope
is a unique feature of IceCube.
It will enable an exploration of the cosmic-ray
spectrum using the combined measurement of the electromagnetic and muon
components of atmospheric showers at energies from below the knee up to
$10^{18}$~eV with unprecedented statistics and energy resolution.
Charge spectra of the DOMs in the 8 IceTop tanks exhibit a muon peak 
at about 240 PE 
indicating a nominal energy deposition of 0.8 MeV/photoelectron, and the
data indicate a uniform response as a function of the deposited energy in the tanks.
The direction of showers has been reconstructed and the expected 
sec($\theta$) dependence is observed.

During the austral summer of December 2005 - January 2006 eight more
strings of 60 DOMs each have been deployed in the ice along with
12 additional surface stations.
 After the new strings are commissioned, the
number of 
operating DOMs with 10 inch PMTs in IceCube will be comparable to the total
number of 8 inch PMTs in the AMANDA modules.  Because of the larger
spacing the effective volume of the combined neutrino detector will
be twice that of AMANDA II and somewhat larger for higher energy events
($> 10$~TeV).  
This will allow
the experiment to start physics analysis, and in a few years the discovery of the
first astrophysical neutrinos is a realistic possibility. With sixteen stations on
the surface the combined detector will also be sensitive to cosmic-rays
up to $10^{17}$~eV. 

The construction is scheduled to advance at a rate of 14 or more strings
and tank stations per year in successive seasons until the detector
reaches its final configuration in 2011.  The plan calls for
4800 In-Ice DOMs on 80 strings, distributed in a volume of 1 km$^3$
in a triangular pattern with 125 m lateral spacing and at depths from 1450 
to 2450 m. The surface array will comprise 160 frozen water
tanks each containing 2 DOMs, which constitute 80 stations  
arranged in pairs associated with each string.
\vskip 2.cm
\section*{Acknowledgments}
{We acknowledge the support of the following agencies: National
Science Foundation--Office of Polar Programs, National Science
Foundation--Physics Division, University of Wisconsin Alumni Research
Foundation, Department of Energy, and National Energy Research
Scientific Computing Center (supported by the Office of Energy
Research of the Department of Energy), the NSF-supported TeraGrid
systems at the San Diego Supercomputer
Center (SDSC), and the National Center for Supercomputing Applications
(NCSA);
Swedish Research Council,
Swedish Polar Research
Secretariat, and Knut and Alice Wallenberg Foundation, Sweden; German
Ministry for Education and Research, Deutsche Forschungsgemeinschaft
(DFG), Germany; Fund for Scientific Research (FNRS-FWO), Flanders
Institute to encourage scientific and technological research in
industry (IWT), and Belgian Federal Office for Scientific, Technical
and Cultural affairs (OSTC); The Netherlands Organisation for Scientific
Research (NWO).}
\newpage
\clearpage

\end{document}